\documentclass[aps,twocolumn, prX,nofootinbib]{revtex4}
\usepackage[usenames, dvipsnames]{color}
\usepackage{graphicx, array}
\usepackage{graphics,epsfig}
\usepackage{amssymb}
\usepackage{amsmath}
\usepackage{ulem}
\usepackage{multirow}
\usepackage{subfigure}
\usepackage{ulem}
\usepackage{gensymb}
\usepackage{rotating}
\usepackage{aas_macros}
\begin{document}

\input{epsf}

\title{Detection of the pairwise kinematic Sunyaev-Zel'dovich effect with BOSS DR11 and the Atacama Cosmology Telescope}
\author {F. De~Bernardis$^1$, S. Aiola$^2$, E.~M. Vavagiakis$^1$, N.~Battaglia$^3$, M.~D.~Niemack$^1$, J.~Beall$^4$, D.~T.~Becker$^4$, J.~R.~Bond$^5$, E.~Calabrese$^6$, H.~Cho$^{4,7}$, K.~Coughlin$^8$, R.~Datta$^8$, M.~Devlin$^9$, J.~Dunkley$^6$, R.~Dunner$^{10}$, S.~Ferraro$^{11,12}$, A.~Fox$^4$, P.~A.~Gallardo$^1$, M.~Halpern$^{13}$, N.~Hand$^{12}$,  M.~Hasselfield$^{14,15}$, S.~W.~Henderson$^1$, J.~C.~Hill$^{16}$, G.~C.~Hilton$^4$, M.~Hilton$^{17}$, A.~D.~Hincks$^{12,18}$, R.~Hlozek$^{19}$, J.~Hubmayr$^4$, K.~Huffenberger$^{20}$, J.~P.~Hughes$^{21}$, K.~D.~Irwin$^{4,7}$, B.~J.~Koopman$^1$ A.~Kosowsky$^2$, D.~Li$^7$, T.~Louis$^{22}$, M.~Lungu$^{9}$, M.~S.~Madhavacheril$^{23}$, L.~Maurin$^{10}$, J.~McMahon$^{8}$, K.~Moodley$^{17}$, S.~Naess$^6$,  F.~Nati$^9$, L.~Newburgh$^{19}$, J.~P.~Nibarger$^4$, L.~A.~Page$^{24}$, B.~Partridge$^{25}$, E.~Schaan$^{3}$, B.~L.~Schmitt$^{9}$, N.~Sehgal$^{23}$, J.~Sievers$^{26,27}$, S.~M.~Simon$^{24}$ D.~N.~Spergel$^{3}$, S.~T.~Staggs$^{24}$, J.~R.~Stevens$^1$, R.~J.~Thornton$^{9,28}$, A.~van~Engelen$^{5}$, J.~Van~Lanen$^4$,  E.~J.~Wollack$^{29}$\vspace{10pt}}
\affiliation{ $^1$ Department of Physics, Cornell University, Ithaca, NY 14853, USA}
\affiliation{ $^2$ Department of Physics and Astronomy, University of Pittsburgh, 3941 OHara Street, Pittsburgh, PA 15260 USA and Pittsburgh Particle Physics, Astrophysics, and Cosmology Center, Pittsburgh, PA 15260 USA}
\affiliation{ $^3$ Department of Astrophysical Sciences, Peyton Hall, Princeton University, Princeton, NJ USA 08544}
\affiliation{ $^4$ National Institute of Standards and Technology, Boulder, CO USA 80305}
\affiliation{ $^5$ CITA, University of Toronto, 60 St. George St., Toronto, ON M5S 3H8, Canada}
\affiliation{ $^6$ Sub-Department of Astrophysics, University of Oxford, Keble Road, Oxford, UK OX1 3RH}
\affiliation{ $^7$ SLAC National Accelerator Laboratory, 2575 Sand Hill Road, Menlo Park, CA 94025}
\affiliation{$^8$ Department of Physics, University of Michigan Ann Arbor, MI 48109, USA}
\affiliation{$^9$ Department of Physics and Astronomy, University of Pennsylvania, 209 South 33rd Street, Philadelphia, PA, USA 19104}
\affiliation{$^{10}$ Instituto de Astrof\'isica and Centro de Astro-Ingenier\'ia, Facultad de F\'isica, Pontificia Universidad Cat\'olica de Chile, Av. Vicu\~na Mackenna 4860, 7820436 Macul, Santiago, Chile}
\affiliation{$^{11}$ Miller Institute for Basic Research in Science, University of California, Berkeley, CA, 94720, USA}
\affiliation{$^{12}$ Astronomy Department, University of California, Berkeley, CA 94720, USA}
\affiliation{$^{13}$ University of British Columbia, Department of Physics and Astronomy, 6224 Agricultural Road, Vancouver BC V6T 1Z1, Canada}
\affiliation{$^{14}$ Department of Astronomy and Astrophysics, The Pennsylvania State University, University Park, PA, 16802, USA}
\affiliation{$^{15}$ Institute for Gravitation and the Cosmos, The Pennsylvania State University, University Park, PA 16802, USA}
\affiliation{$^{16}$ Dept. of Astronomy, Pupin Hall, Columbia University, New York, NY, USA 10027}
\affiliation{$^{17}$ School of Mathematics, Statistics and Computer Science, University of KwaZulu-Natal, Durban, 4041, South Africa}
\affiliation{$^{18}$ Department of Physics, University of Rome La Sapienza, Piazzale Aldo Moro 5, I-00185 Rome, Italy}
\affiliation{$^{19}$ Dunlap Institute, University of Toronto, 50 St. George St., Toronto ON M5S3H4}
\affiliation{$^{20}$ Florida State University, Tallahassee, FL 32306, USA}
\affiliation{$^{21}$ Department of Physics and Astronomy, Rutgers University, 136 Frelinghuysen Road, Piscataway, NJ 08854-8019}
\affiliation{$^{22}$ UPMC Univ Paris 06, UMR7095, Institut dAstrophysique de Paris, F-75014, Paris, France}
\affiliation{$^{23}$ Stony Brook University, Stony Brook, NY 11794}
\affiliation{$^{24}$ Joseph Henry Laboratories of Physics, Jadwin Hall, Princeton University, Princeton, NJ 08544, USA}
\affiliation{$^{25}$ Department of Physics and Astronomy, Haverford College, Haverford, PA, USA 19041}
\affiliation{$^{26}$ School of Chemistry and Physics, University of KwaZulu-Natal, Durban, 4041, South Africa}
\affiliation{$^{27}$ National Institute for Theoretical Physics (NITheP), KZN node, Durban, South Africa}
\affiliation{$^{28}$ Department of Physics, West Chester University of Pennsylvania, West Chester, PA 19383, USA}
\affiliation{$^{29}$ NASA Goddard Space Flight Center, 8800 Greenbelt Road, Greenbelt, Maryland 20771, USA}

\label{firstpage}

\begin{abstract}
We present a new measurement of the kinematic Sunyaev-Zel'dovich effect
using data from the Atacama Cosmology Telescope (ACT) and the Baryon
Oscillation Spectroscopic Survey (BOSS). Using 600 square degrees of
overlapping sky area, we evaluate the mean pairwise baryon momentum
associated with the positions of 50,000 bright galaxies in the BOSS
DR11 Large Scale Structure catalog. A non-zero signal arises from
the large-scale motions of halos containing the sample galaxies. The
data fits an analytical signal model well, with the optical depth to
microwave photon scattering as a free parameter determining the
overall signal amplitude. We estimate the covariance matrix of the
mean pairwise momentum as a function of galaxy separation, using
microwave sky simulations, jackknife evaluation, and bootstrap
estimates. The most conservative simulation-based errors give
signal-to-noise estimates between 3.6 and 4.1 for varying galaxy
luminosity cuts. We discuss how the other error determinations can
lead to higher signal-to-noise values, and consider
the impact of several possible systematic errors. Estimates of the
optical depth from the average thermal Sunyaev-Zel'dovich signal at the
sample galaxy positions are broadly consistent with those obtained
from the mean pairwise momentum signal.
\end{abstract}

\maketitle

%%%%%%%%%%%%%%%%%%%%%%%%%%%%%%%%%%%%%%%%%%%%%%
\section{Introduction}
The kinematic Sunyaev-Zel'dovich effect \cite{1972CoASP...4..173S} (kSZ) is the only known way to directly measure the peculiar velocities of objects at cosmological distances.
A moving galaxy cluster containing ionized gas creates a near-blackbody
spectral distortion in the microwave background radiation passing through it,
with an amplitude proportional to both the total gas mass and the line-of-sight
velocity component, but independent of the gas temperature. The kSZ effect is hence a valuable source of information for cosmology, allowing tests of dark energy and gravity between megaparsec and gigaparsec scales \cite{Bhattacharya:2007sk,Mueller:2014dba,Mueller:2014nsa}.

For large clusters ($M>10^{14} M_{\odot}$), the kSZ signal is more than an order of magnitude smaller than the typical thermal SZ \cite{1970Ap&SS...7....3S} (tSZ)
spectral distortion at most wavelengths. Detecting the velocity of
individual clusters requires measurements at multiple frequencies and
high precision along with models of the intracluster medium. To date,
the only claimed detection of a peculiar velocity for a single object
comes from Bolocam observations of the cluster MACS J0717.5+3745,
giving a high peculiar velocity for a particular subcluster $v = 3450\pm900$ km/s \cite{Sayers:2013ona}.
This analysis modeled the tSZ signal from X-ray data and then
jointly fitted the thermal and kinematic SZ signals. A more recent analysis \cite{2016arXiv160607721A} mapped the kSZ effect
for a single cluster using data from the New IRAM KID Arrays \cite{2011ApJS..194...24M} detecting the dipolar signature associated with two merging subclusters.

For lower-mass clusters, the kSZ and tSZ signals are comparable but 
both signals are small compared to the noise level in current microwave
background maps. The Atacama Cosmology Telescope and SDSS collaborations made the
first statistical detection of the kSZ signal by estimating the mean
pairwise cluster momentum from a sample of clusters identified by their
bright central galaxies in the Sloan Digital Sky Survey (SDSS) \cite{2012PhRvL.109d1101H}. 
The nonzero mean pairwise velocity of galaxy clusters reflects
the slight tendency of any pair of clusters to be
moving towards each other, due to the attractive force of gravity.
This statistic is also advantageous because it is a linear difference of
measured sky temperatures at the positions of clusters, and most other
signals, like tSZ and dust emission, average out. Recently, detections using
the same estimator have been reported by the Planck
collaboration using galaxies from SDSS \cite{Ade:2015lza} and the South
Pole Telescope collaboration using galaxies from the Dark Energy Survey \cite{2016arXiv160303904S}. The former work has been used in \cite{PhysRevLett.115.191301} to measure the amount of missing baryons. The latter analysis reported a $4.2\sigma$ detection, showing for the first time 
that the pairwise kSZ signal can be extracted using photometric data once the redshift uncertainty is properly taken into account.

%Statistical detection of the kinematic SZ effect has also been 
%achieved with a different
%technique: first, construct a velocity template from the large-scale
%density field from galaxy surveys assuming the continuity equation,
%and then cross-correlate this velocity field with the microwave
%background temperature map \cite{Schaan:2015uaa, Ade:2015lza}. These
%detections are consistent with those from mean pairwise estimates.
%although require more complex modeling and analysis.

A statistical detection of the kSZ effect has also been achieved with a different technique: a velocity template is constructed from the BOSS large-scale density field assuming the continuity equation, then the velocity template is cross-correlated with the CMB temperature map \cite{Schaan:2015uaa, Ade:2015lza}. Schaan et al. (2016) \cite{Schaan:2015uaa} in particular measured the amplitude of the kSZ signal as a function of the angular radius around the clusters and reported $2.9\sigma$ and $3.3\sigma$ evidence of the kSZ using two different velocity reconstruction methods. The ACTPol \cite{2016arXiv160506569T} CMB map used by Schaan et al. (2016) is similar to the one used in this work. It was combined with the CMASS galaxy catalog from BOSS DR10 for the analysis. The Schaan et al. (2016) approach is based on converting galaxy stellar mass estimates to total masses for the host halos and then to an optical depth by using the cosmological baryon abundance. It provides another potential probe of the fraction of free electrons and the baryon profile of galaxy clusters. These detections are consistent with the pairwise measurements presented here and offer a complementary tool to investigate the physics of galaxy clusters and to understand potential systematic effects.

A different way to detect the kSZ effect that involves squaring the CMB anisotropy maps \cite{Ferraro:2016ymw} has been implemented by Hill et al. \cite{Hill:2016dta} using publicly available data. Specifically, foreground-cleaned CMB temperature maps constructed from multi-frequency Planck and WMAP data were filtered, squared, and cross-correlated with galaxy measurements from the Wide-field Infrared Survey Explorer (WISE) \cite{2010AJ....140.1868W}, which yielded a 3.8--4.5 $\sigma$ kSZ detection, depending on the galaxy bias constraints. The advantage of this method is that it does not require redshift estimates for individual clusters, which allows use of photometric data without treating redshift uncertainty.

Here we report a detection of the pairwise kSZ signal using
data from two-year maps of the Atacama Cosmology Telescope Polarimeter
(ACTPol) experiment, combined with galaxy positions and redshifts from
the Baryon Oscillation Spectroscopic Survey (BOSS). The deep portion of
the ACTPol maps provides around 600 deg$^2$ of overlap with the DR11
release of BOSS, which we use to obtain a $4.1\sigma$ detection of
the mean pairwise kSZ effect. This is an improvement over the initial
detection in Hand et al.\ (2012), H12 hereafter, and is comparable to the
significance reported in \cite{2016arXiv160303904S}, which uses a deeper optical data set but
lacks spectroscopic redshifts. 

In Section \ref{sec:data} we describe the ACT, ACTPol, and BOSS data used for this analysis. Section \ref{sec:analysis} summarizes the pairwise statistic approach, the map filtering method used and the model fitting technique that we use to estimate the cluster optical depths and to quantify the significance of the detection. In Section \ref{sec:covariance} we describe three different approaches to estimate the covariance matrix of the data: simulated CMB maps, bootstrap and jackknife. In Section \ref{sec:results} we show the results of our analysis and in Section \ref{sec:comparison} compare with the previous ACT results reported by H12, discussing the differences in the map filtering approach and in the covariance matrix estimation method. Finally, in Section VI we use the same dataset to extract the tSZ component and use the results of hydrodynamical simulations to obtain an estimate of the optical depth independent of the one obtained by the kSZ. We show that measurements of the tSZ combined with hydrodynamical simulations can be used to estimate the average optical depth and thereby convert the pairwise momentum measurement into a pairwise velocity.

\section{Data}\label{sec:data}
\subsection{ACTPol data}
\begin{figure*}[ht]
\begin{center}  
\hspace*{-1cm}
\includegraphics[width=18cm]{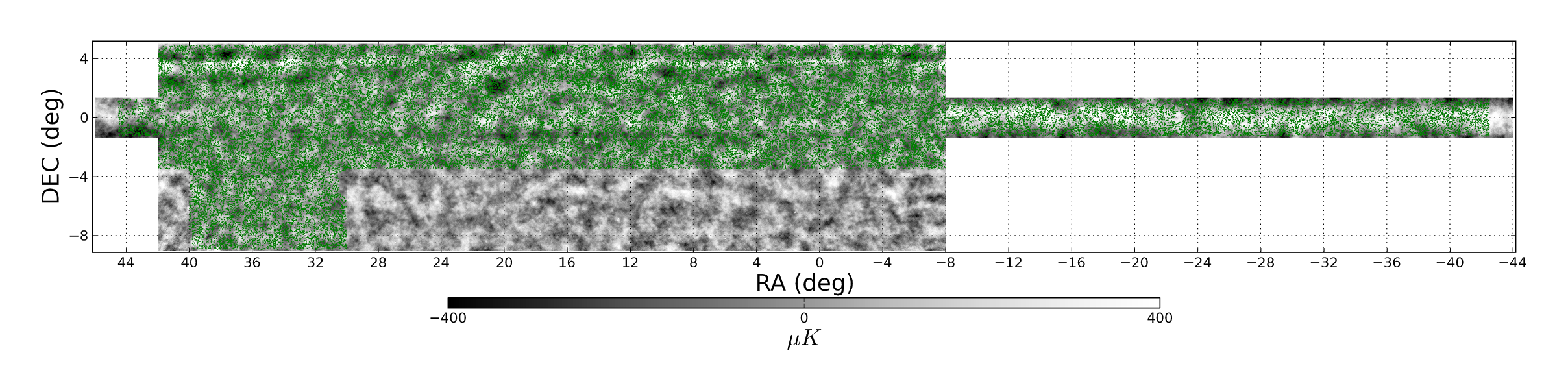}
\caption{ACTPol map used for this analysis, showing the overlap with the $67938$ DR11 sources (green dots). The long strip across the field is the region observed by ACT that was used for the first measurement of the pairwise kSZ effect in \cite{2012PhRvL.109d1101H}.}\label{fig:map}
\end{center}
\end{figure*}

The CMB data used for this analysis are the combination of sky maps at 148 GHz from two seasons of ACT observations and two seasons of nighttime observations with the ACTPol receiver. The ACT data and the ACTPol data used in this work cover different areas of the sky, with partial overlap. The ACT data used here are presented in \cite{Das:2013zf}, while the data for the first season (2013) of ACTPol are presented in \cite{Naess:2014wtr}. This coadded map includes the deep patches labeled D5 and D6, centered respectively at right ascensions (RA) $-5\degree$ and $35\degree$ near the celestial equators and the region encompassing D5 and D6, called deep56, which was observed in 2014 by ACTPol \cite{fdb_spie}. This map is the deepest patch of the sky observed by ACTPol and covers about $700$ deg$^2$ with a white noise level that ranges from 10 (for the deepest regions) to 20 $\mu$K$\cdot$arcmin. Fig. \ref{fig:map} shows the ACTPol data used in this paper with the sources from the BOSS catalog, which overlaps with about 600 deg$^2$ of the coadded map.
 
\subsection{BOSS-SDSS data}
\begin{figure}[!h]
\begin{center}
\hspace*{-1.5cm}
\includegraphics[width=11cm]{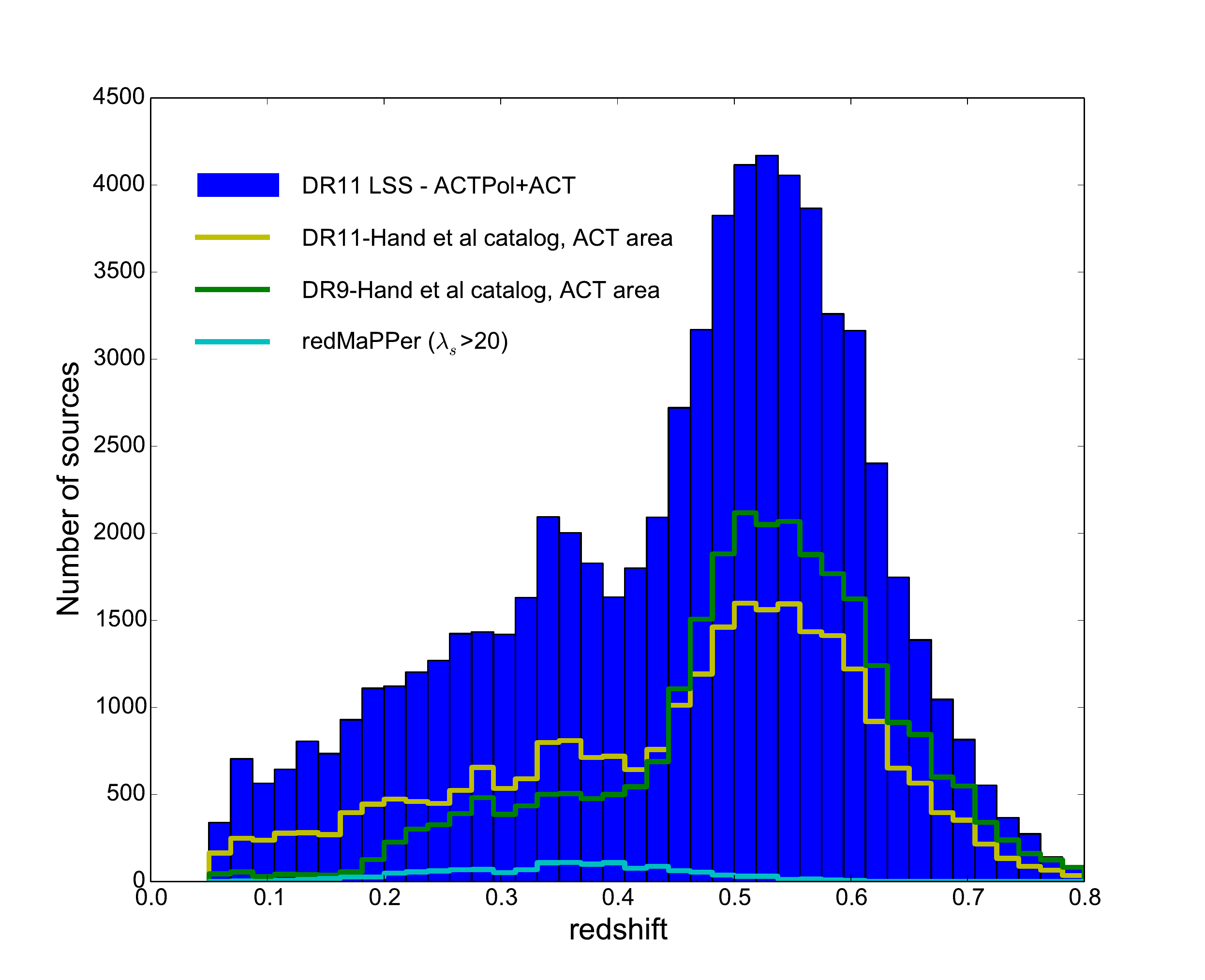}
\caption{Comparison between the DR9 selection used in \cite{2012PhRvL.109d1101H} (green) for the 220 sq. deg ACT region and the DR11 LSS catalog overlapping the same area (yellow). The two catalogs contain 27291 and 26357 galaxies respectively. The blue histogram shows the DR11 redshift distribution in the coadded ACT+ACTPol map that has about 600 sq deg of overlap with the DR11 catalog. The light blue is the distribution for the redMaPPer sources with richness $\lambda_s>20$.}
\label{fig:catalogs}
\end{center}
\end{figure}

We use the public Large Scale Structure (LSS) DR11 catalog\footnote{\url{http://data.sdss3.org/sas/dr11/boss/lss/}} 
from the BOSS survey \cite{2013AJ....145...10D}. We measure a temperature signal from our CMB maps in the direction of these objects and assume that the most luminous of these galaxies coincide with the center of clusters. BOSS sources within 5 arcmin radius from point sources with flux $>15$ mJy are discarded.
In \cite{2012PhRvL.109d1101H}, a 220 deg$^2$ ACT map was combined with a selection of the DR9 sources. For comparison we repeat our analysis using the DR9 selection used in that paper. These two catalogs contain roughly the same number of objects in the ACT-only map, i.e. 27291 for the DR9 selection and 26357 for the DR11 LSS catalog, although the redshift distributions are significantly different (see Fig. \ref{fig:catalogs}), with the DR9 selection having more high-redshift objects. The redshifts range from $0.05$ to $0.8$ with an average reshift of $0.48$ for the DR11 catalog and $0.56$ for the DR9 selection. 

For the coadded map we find 67938 sources in the same redshift range from the DR11 catalog. The luminosities of these sources are calculated based on their r-band Petrosian de-reddened magnitudes and applying a K-correction using the \url{k_correct} software \cite{Blanton:2006kt}. The luminosities range from  $1.5\times 10^{8}L_{\odot}$ to $1.25\times 10^{12}L_{\odot}$.

We also analyze the redMaPPer \cite{Rykoff:2013ovv,Rozo:2013vja,Rozo:2014zva}  SDSS DR8 catalog of galaxy clusters. The redMaPPer algorithm provides a more precise determination of the center of the cluster, at the expense of a smaller sample of objects and with some of the redshifts estimated photometrically.
We find $31600$ clusters in the coadded ACTPol-ACT map, of which only $2242$ have a richness $\lambda_s>20$ after including a correction factor accounting for masked clusters and incompleteness. See Section \ref{sec:redmapper} for more details.

\section{Analysis}\label{sec:analysis}
We implement a pairwise statistic following the approach successfully employed on ACT and BOSS DR9 data in \cite{2012PhRvL.109d1101H} for the first kSZ measurement, although our analysis differs significantly from \cite{2012PhRvL.109d1101H}. As described below, we use aperture photometry to filter the CMB maps as opposed to the matched filter used in \cite{2012PhRvL.109d1101H}, and we compare three different methods for uncertainty estimation.
\subsection {Pairwise momentum estimator and kSZ effect}
A well-known prediction of linear perturbation theory is that the mean pairwise momentum of galaxy clusters as a function of their comoving separation is negative for small separations \cite{1991ApJ...374L...1H,1994MNRAS.271..976N,Juszkiewicz:1998xf,Sheth:2000ff}, implying that galaxy clusters move towards each other on average. The absolute value of the pairwise momentum peaks around separations of 30 Mpc. Direct measurements of the pairwise momenta of clusters would require knowledge of their 3-dimensional momentum which is not easily measurable.  The mean pairwise momentum $p$ can be still estimated from the line-of-sight component of the momenta of a sample of clusters through the estimator \cite{Ferreira:1998id}:
\begin{equation}\label{estimator}
p_{\rm est}(r) = \frac{\sum_{i<j}(\mathbf{p}_i\cdot \mathbf{r}_i- \mathbf{p}_j\cdot \mathbf{r}_j)c_{ij}}{\sum_{i<j}c_{ij}^2}
\end{equation}
where $\mathbf{r_i}$ is the comoving distance to the $i^{th}$ cluster calculated from the redshift of the cluster assuming a fiducial value for the cosmological parameters. Here, $r$ is the comoving separation between clusters $i$ and $j$, $r = |\mathbf{r}_{ij}|=|\mathbf{r}_i-\mathbf{r}_j|$ and the pairwise estimator at separation $r$ is obtained by summing over all the pairs satisfying $|\mathbf{r}_{ij}| = r$ in equation (\ref{estimator}). The factor $c_{ij}$ accounts for the alignment of a pair of clusters $i$ and $j$ along the line of sight:
\begin{equation}
c_{ij} = \mathbf{r}_{ij}\cdot\frac{\mathbf{r}_i+\mathbf{r}_j}{2} = \frac{(r_i-r_j)(1+\cos \theta)}{2(r_i^2+r_j^2-2r_ir_j \cos\theta)}
\end{equation}
with $\theta$ being the angular separation between two clusters. The line-of-sight momentum of a given cluster is related to the measured kinematic SZ signal via a simple direct proportionality: $\Delta T_{kSZ,i} \propto -\mathbf{p}_i\cdot\mathbf{r}_i$, where the multiplicative factors depend on the cluster properties (density profile and hence its angular extent in the sky) and on the pixel scale and beam of the CMB experiment. Extracting the kSZ signal from a single cluster is a challenging task, as the kSZ temperature fluctuation can be orders of magnitude lower than other effects, such as the tSZ or the infrared emission. The advantage of (\ref{estimator}) is that, being a differential measurement, for a large enough sample of clusters any signal that does not depend on the comoving separation should average to zero. The primary source of contamination can be caused by redshift (and hence comoving distance) dependent effects that could mimic a pairwise signal dependent on the separation for every pair of sources having significantly different redshifts. Following \cite{2012PhRvL.109d1101H}  we account for these effects by removing a Gaussian weighted average temperature signal of the sources as function of redshift:
\begin{equation}\label{corrz}
T_{z}(z) = \frac{\sum_i \Delta T_i(z_i)\exp{[-(z-z_i)^2/\sigma_z^2]}}{\sum_i\exp{[-(z-z_i)^2/\sigma_z^2]}}
\end{equation}
where we use $\sigma_z = 0.01$. We have verified that this correction is only weakly dependent on $\sigma_z$, whose value does not significantly affect the signal-to-noise ratio. We can hence rewrite equation (\ref{estimator}) as:

\begin{eqnarray}\nonumber\label{estimator_t} 
p_{\rm est}(r) =&&\\
&-\frac{\sum_{i<j}[(\Delta T_i-T_{z}(z_i))-( \Delta T_j-T_{z}(z_j))]c_{ij}}{\sum_{i<j}c_{ij}^2}.&
\end{eqnarray}
%\begin{equation}
%\begin{split}
%p_{\rm est}=\\ 
%\quad-\frac{\sum_{i<j}[(\Delta T_i-T_{z}(z_i))-( \Delta T_j-T_{z}(z_j))]c_{ij}}{\sum_{i<j}c_{ij}^2}
%\end{split}
%\end{equation}
We apply this estimator to our catalogs, dividing the comoving separations range into bins of width $10$ Mpc and extending the calculation up to $400$ Mpc separations.

\subsection{Filtering CMB maps}\label{sec:filt}
In order to associate a temperature signal to a given source we use an aperture photometry (AP) filter. For each source we select a circular disk of a given angular radius (aperture) $\alpha$, take the average value of the pixels belonging to the disk $T_{\rm disk}$, and subtract from this average the average value of the pixels ($T_{\rm ring}$) in a ring of radii [$\alpha$, $\alpha\sqrt 2$]. The temperature associated with a source $i$ will be $T_i = T_{\rm disk,i}-T_{\rm ring,i}$. This AP filter has been applied for a similar pairwise statistic analysis in \cite{Ade:2015lza} and offers the advantage of being independent of assumptions about the shape of the density profile of the cluster, unlike the matched filter approaches used in \cite{2012PhRvL.109d1101H,2016arXiv160303904S}. As discussed in \cite{Ade:2013opi} the $\sqrt 2$ factor for the outer radius of the ring is optimized to remove local fluctuations on scales just above the scale of the inner radius $\alpha$ without introducing noise by having a too small ring area. Even though it does not involve assumptions about the shape of the density profile, the AP filter does assume that the cluster is contained within the aperture. Using small apertures would lead to a subtraction of the signal while large apertures would mix the signal with the background CMB and noise. For a given map noise, angular resolution and cluster profile it is straightforward to show that there exists an optimal aperture which maximizes the signal-to-noise ratio for each cluster.

\subsection{Model fitting}\label{sec:modelfitting}
To estimate the significance of our detection we use the analytic prediction of linear perturbation theory for the pairwise velocity \cite{1991ApJ...374L...1H,1994MNRAS.271..976N,Juszkiewicz:1998xf,Sheth:2000ff,Bhattacharya:2007sk,Mueller:2014dba,Mueller:2014nsa}, rescaled by a negative factor $-\bar{\tau}T_{CMB}/c$, where $\bar{\tau}$ is the free parameter of our fit and can be interpreted as an average optical depth of the cluster sample used for the pairwise momentum estimator and averaged over the aperture of the filter. This average value is hence related only to the number of free electrons within the aperture of the filter. Clusters at lower redshifts in particular can have angular sizes larger than the filter aperture. In this case the AP filter approach is likely to underestimate the optical depth of the cluster. However, we have verified that scaling the aperture of the filter with the cluster redshift affects the results by $\lesssim 0.05\sigma$. 

It is known that this linear model fails at lower separations ($r<20$ Mpc) because of redshift space distortions and higher order nonlinear effects that cause the real pairwise momentum to become positive rather than approaching zero from negative values as predicted by the linear model \cite{2012JCAP...11..009V,Okumura:2013zva,Sugiyama:2015dsa}. For this reason we exclude the first two bins when fitting to our linear model template. We include nonlinear redshift space distortion effects following the approach of \cite{2012JCAP...11..009V,Okumura:2013zva} and repeat the fit to the data extending it to the entire range of comoving separation.

The minimum mass for the sources used to calculate the pairwise estimator is estimated with a halo model \cite{Cooray:2002dia} assuming the empirical galaxy bias-luminosity relation of \cite{2004ApJ...606..702T,2005ApJ...630....1Z} and the bias-halo mass relation based on numerical simulations from \cite{2010ApJ...724..878T}. We use the same mass cut and average redshift of these sources when calculating the theory curve template. To calculate the theory templates and to reconstruct the comoving separations from the source redshifts we assume the Planck cosmology for a flat universe \cite{2015arXiv150201589P}: $\Omega_bh^2 = 0.02225$, $\Omega_ch^2 =0.1198$, $H_0 = 67.3$ km/s/Mpc, $\sigma_8 = 0.83$, $n_s = 0.964$.

%\subsection{Null tests}
\section{Covariance matrix}\label{sec:covariance}
We implement three different methods to estimate the covariance matrix. The first approach uses four hundred mock CMB maps including anisotropic instrumental noise and atmospheric noise. We create several realizations by running our estimator on each one of these simulated maps using the coordinates of the sources in the catalog to estimate the temperature signal and their redshift to calculate the geometrical weights in equation (\ref{estimator_t}). The $c_{ij}$ factors are hence the same for all the realizations while the temperature values associated with each position change from one realization to the other. We then use these realizations to calculate the covariance matrix for 40 comoving separation bins in the range $0-400$ Mpc. The main advantage of this method is that the mock maps are independent of each other and the estimated covariance matrix properly includes the cosmic variance contribution from the CMB. Moreover using mock maps of CMB and noise provides a useful null test by taking the average of many null realizations. We consider this the preferred method for estimating the covariance matrix, with the drawback of being more time consuming.

The second method consists of creating bootstrap realizations of the pairwise momentum estimator by resampling the temperature differences (including the redshift correction) that enter in the estimator (\ref{estimator_t}). For each realization the temperature difference values in a separation bin are replaced with values randomly selected from the distribution of all the temperature difference values. Bootstrap resampling the temperature differences reduces the risk of a redshift dependent contamination that might bias the covariance matrix if the correction in (\ref{corrz}) fails to account for these effects. Moreover, it does not require creating CMB simulations and can be implemented starting from the data sample itself without external inputs. On the other hand, the temperature difference values are correlated among the various comoving bins because each source can contribute to different bins multiple times. As a consequence, shuffling the values of temperature differences can significantly alter the estimated covariance matrix in a complicated way. As expected we find that the bootstrap tends to systematically underestimate the error bars compared to mock maps especially at separations larger than $\gtrsim 150$ Mpc, while for $r\lesssim 150$ Mpc the differences are smaller, on the order of $\sim10 \%$ and does not recover the full correlation between comoving separation bins.

The third method is a jackknife approach similar to the one implemented by \cite{2016arXiv160303904S}. We split the sample of sources into $N$ smaller subsamples and remove one subsample at a time, running the pairwise estimator on the union of the remaining $N-1$ subsamples, obtaining $N$ realizations. The fact that these realizations are not independent can be accounted for by a $N-1$ factor when calculating the covariance:
\begin{equation}\label{cov_jk}
  C_{mn} = \frac{N-1}{N}\sum_{k=1}^{N}\left(p^k_m-\bar{p}_m\right)\left(p^k_n-\bar{p}_n\right)
\end{equation}
where $m$ and $n$ are the indices for the comoving separation bins and $p^k_{m,n}$ is the pairwise estimator value in the bins $m$ and $n$ for the realization $k$. The inverse covariance matrix in (\ref{cov_jk}) is generally a biased estimator of the true inverse covariance matrix. It is possible to correct for this bias by multiplying $C_{m,n}^{-1}$ by (${N-K-1})/N-1$ \cite{Hartlap:2006kj}, where $K$ is the number of comoving separation bins. Similar to the bootstrap approach, the jackknife method has the advantage of being based on the dataset itself and of not requiring external maps. On the other hand, compared to the bootstrap, it is easier to account for different realizations not being independent by using (\ref{cov_jk}). Moreover, since there is no shuffling involved, but only subsamples removed from the main cluster sample, the covariance between bins is recovered with better accuracy than it is with the bootstrap method. We find that this jackknife technique provides a detection consistent with the one estimated from mock maps, with differences $\lesssim 0.5\sigma$.

We perform a full comparison of the detection significance achieved by using different error estimation approaches. Table \ref{lcut} shows the signal-to-noise ratio estimated with the bootstrap and jackknife methods. It can be seen that the bootstrap of temperature differences overestimates the signal-to-noise compared to simulated maps when considering separations above $150$ Mpc. For example for the case $L> 7.9 \times 10^{10} L_{\odot}$ the S/N ratio from bootstrap is $8.8$ while simulated maps provide $4.1$. The sharp change in S/N results largely from ignoring the correlation between comoving bins when the bootstrap covariance matrix is constructed. We find a maximum correlation value of around $0.2$ even at the larger separations, which is significantly lower than the expected correlation from simulated maps with a maximum of $0.55$ for $r<150$ Mpc. This can be confirmed by using only the diagonal part of the mock map covariance matrix. In that case we find a S/N of $7$ from simulations, closer to the $8.8$ from bootstrap. %As discussed in the previous section another source of tension between the bootstrap and the simulations approach is the lack of cosmic variance in the former.

The jackknife approach, on the other hand, produces results more consistent with the simulated maps. As discussed above, this is because with the jackknife resampling we are only removing subsamples from the main cluster sample, which better preserves the correlation between bins. Moreover it is straightforward to account for the realizations not being independent with the $N-1$ factor in equation (\ref{cov_jk}). %Figure \ref{fig:errors_comp} shows the ratio of the error bars estimates (the diagonal part of the covariance matrix) to those obtained from mock simulated maps. It can be seen that the error bars from bootstrap are heavily underestimated, while those from the jackknife are closer to
For the jackknife covariance matrix we find a maximum correlation value of $0.26$ for bins $r<150$ Mpc, which is the range of separations that contributes to most of the S/N. This is a factor of $2$ lower than the correlation found from simulated maps, suggesting that simulations are still the most conservative approach. Despite this difference, the S/N for the jackknife method is close to the one found with simulations. To check the convergence of the jackknife algorithm we have varied $N$ and looked for the value that provided a stable significance against variations of $N$. We used the same approach for the other cases in Table \ref{lcut} and looked for the $N$ that provided a stable S/N.

\subsection{Contributions to the covariance matrix}
The covariance matrix for the mean pairwise momentum includes contributions from several terms: cosmic variance, primordial CMB and instrumental and atmospheric noise. In addition to these terms, the thermal SZ effect, particularly for clusters with masses $M\gtrsim 10^{14}M_{\odot}$,  can potentially contaminate the kSZ statistics as found by Soergel et al. in \citep{2016arXiv160303904S}. The galaxies used in this work are associated with less massive halos. To verify that the tSZ is not a significant contribution to the noise, we have removed the most luminous sources from our sample, corresponding to masses $M\gtrsim 10^{14}M_{\odot}$ as based on the halo model. For this mass cut we have removed only 237 galaxies from our sample and we have found that the significance of the kSZ pairwise momentum detection is not affected by this cut.

In order to estimate the relative contribution of the various noise sources to the covariance matrix, we have produced two additional sets of simulated maps: simulated maps of noise-free primary CMB, and simulated maps of instrumental and atmospheric noise, not including primary CMB anisotropies. We run our estimator on these simulated maps 400 times for each set and calculate the covariance matrix for each of these sets. We have found that the instrumental plus atmospheric noise contributes 80$\%$ of the total covariance matrix, while the primary CMB represents the remaining $20\%$ of the uncertainty. 

It should be noted that none of the methods mentioned above, including the CMB simulations approach, accounts for the cosmic variance of the velocity field. The cosmic variance contribution can be estimated analytically, for example by following the calculations in \cite{Bhattacharya:2007sk,Mueller:2014dba,Mueller:2014nsa}. We have found that, despite the relatively small sky area ($f_{sky}\simeq0.013$), the redshift range is large enough (z = 0.1-0.8) to provide a significant cosmological volume. We estimate that the cosmic variance contribution is $1\%$ or less (depending on the separations) of the current uncertainty and is not expected to provide a significant contribution to the noise. Future more sensitive surveys might have to account for this effect, depending on the cosmological volume used for the kSZ pairwise statistic. For example, depending on the selection of redshift bins, surveys attempting to reconstruct the pairwise signal as a function of redshift will need to apply the estimator to smaller volumes. The cosmic variance can become significant in those cases.
\section{Results}\label{sec:results}
\subsection{ACTPol+ACT and DR11}
\begin{figure}[h!]
\begin{center}
\includegraphics[width=9cm]{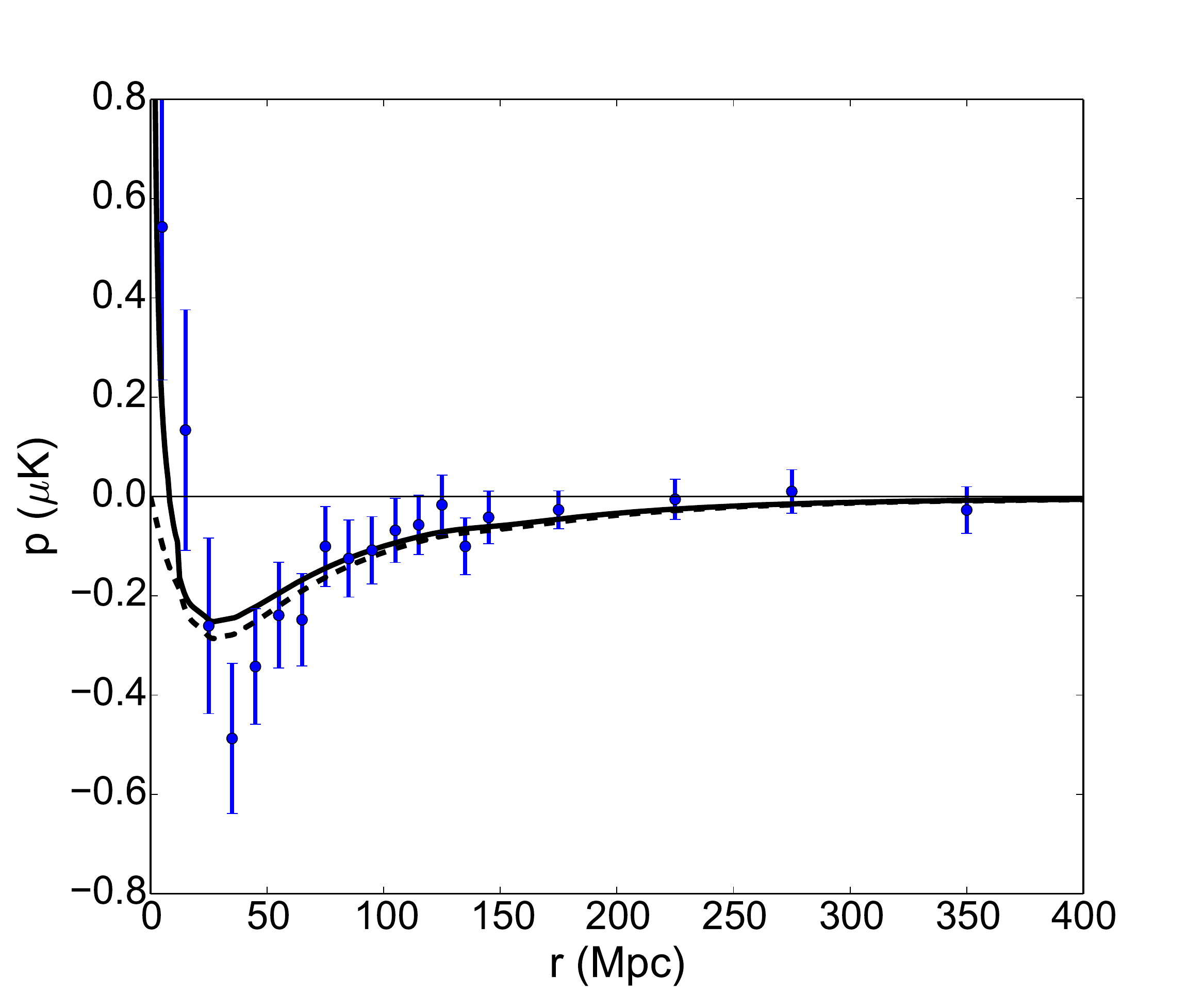}
\includegraphics[width=10cm]{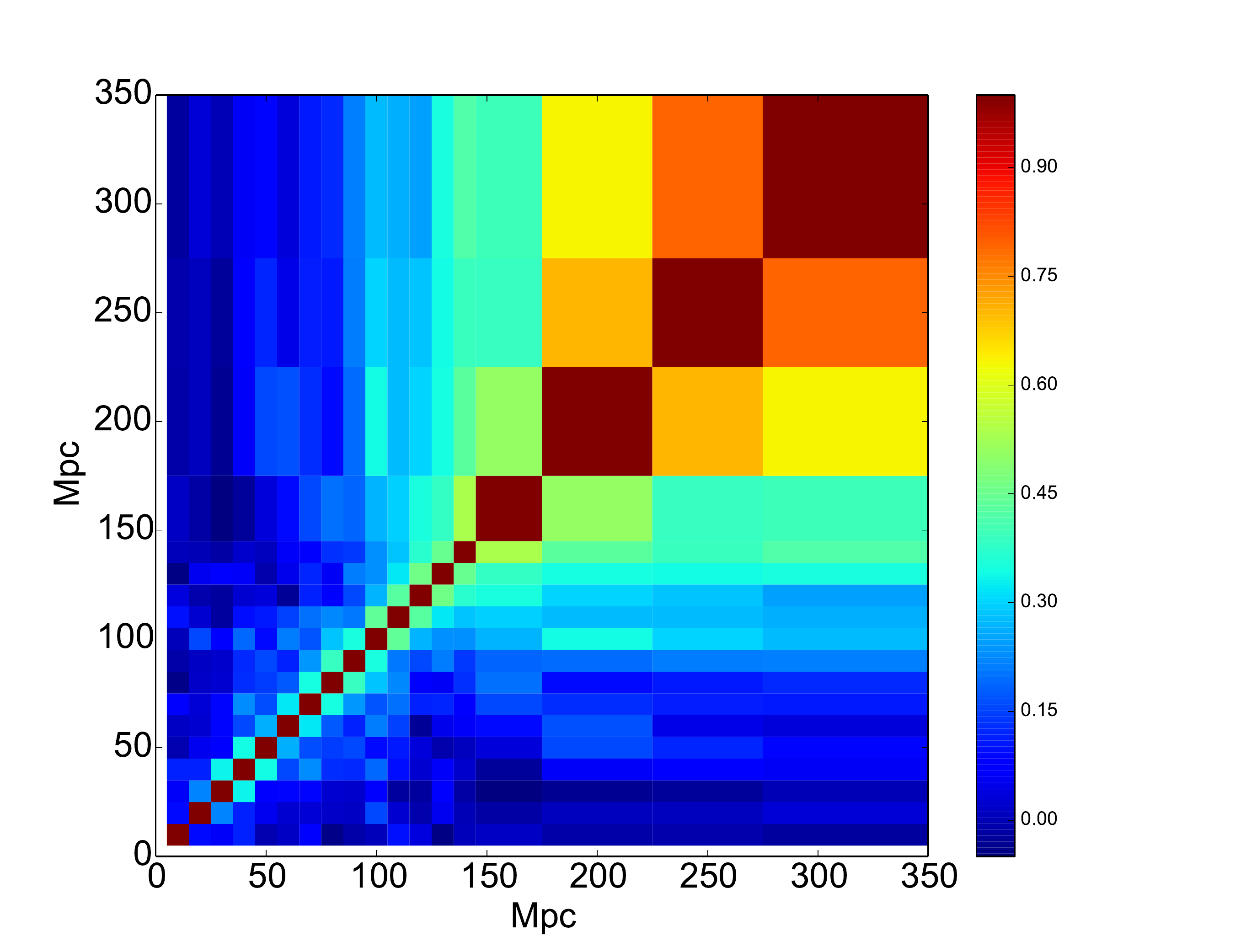}
\caption{Top: Mean pairwise momentum estimator and best fit model for a selection of 20000 objects ($L> 7.9 \times 10^{10} L_{\odot}$) from the DR11 Large Scale Structure catalog. Based on the halo model these luminosities correspond to masses $M>4.6\times10^{13}M_{\odot}$. The filter aperture is $1.8$ arcmin. The best fit average optical depth is $\bar{\tau} = (1.46\pm 0.36)\times 10^{-4}$ when fitting the linear model (dashed line) to separations $r>20$ Mpc, providing a $4.1 \sigma$ evidence of the kSZ signal in the coadded ACTPol-ACT map. The solid line shows the model prediction including nonlinear redshift space corrections. The covariance matrix is estimated from four hundred CMB plus noise realizations of the same area. Bottom: binned correlation matrix for the comoving separation bins of the estimator shown in the top plot, estimated from mock CMB maps. Note the strong correlation at large separations even with large bins.}
%Mean pairwise momentum for a matched filter with fwhm = $1.4^{\prime}$ and an aperture photmetry filter with $2.25^{\prime}$ aperture, for the same selection of sources and CMB data used in \cite{2012PhRvL.109d1101H}.}\vspace{-.2in}
\label{fig:coaddDR11}
\end{center}
\end{figure}

We explore the dependence of the signal on the minimum luminosity of the sample in eight bins in the range $5.3\times 10^{10} L_{\odot} - 11.6\times 10^{10} L_{\odot}$, corresponding to the brightest 50000 and 5000 sources respectively. The bins are chosen by using the 5000 most luminous objects, then increasing the number of objects in steps of 5000 (and steps of 10000 for the largest bins). We study the significance of the detection as function of the luminosity cut to avoid possibe look-elsewhere effects, where one can find an arbitrarily high signal-to-noise ratio by looking for the highest fluctuation in the S/N as a function of the luminosity cut. The mean pairwise momentum estimator and the best fit model are shown in Fig. \ref{fig:coaddDR11} for a selection of 20000 objects ($L> 7.9 \times 10^{10} L_{\odot}$) from the DR11 Large Scale Structure catalog and a filter aperture of  $1.8$ arcmin. This aperture is consistent with the radius required to efficiently remove the CMB plus map noise from the ACTPol map, which can be estimated analytically (see for example \cite{Ferraro:2014cha}). The signal-to-noise ratio is stable for small variations around this aperture. As mentioned above, significantly smaller apertures would subtract the signal while larger apertures would not remove the background efficiently.

The best fit average optical depth for this case is $\bar{\tau} = (1.46\pm 0.36)\times 10^{-4}$, providing $4.1\sigma$ evidence of the kSZ signal in the coadded ACTPol-ACT map as determined from 400 simulations. The model is a good fit to the data with a best fit $\chi^2$ of 23 for 37 dof, including correlations between bins. 
Error bars are estimated by running our estimator on 400 simulations of the CMB sky including noise. We estimate the detection significance and the best fit average optical depth with a $\chi^2$ statistic including the full covariance matrix of the estimator. In Fig. \ref{fig:coaddDR11} we also show the correlation matrix estimated from the mock maps. We find a significant correlation between comoving separation bins for the larger separation bins. For this reason we bin differently the large separation data points when plotting the estimator and the covariance matrix but conduct the analysis on the 40 equally spaced bins with size 10 Mpc. We have verified that the significance of the detection is not affected by the binning. The increased correlation at the largest separations is expected from analytical calculations (see for example \cite{Mueller:2014dba}) because the number of sources in common between pairs belonging to neighboring bins increases at larger separations. The correlation implies that most of the contribution to the S/N is determined by the lower comoving separation bins. We find that excluding bins above 150 Mpc changes the significance of the measurement by $\lesssim 0.1\sigma$. Considering only the diagonal part of the covariance matrix overestimates the signal to noise, yielding up to $S/N\simeq7$, implying that even correlations at the $20-30\%$ level like those observed at $r<150$ Mpc have a significant impact on the overall signal-to-noise.

We have excluded from the fit the two data points at comoving separations $\lesssim 20$ Mpc. Scales smaller than those are affected by nonlinear effects and redshift space distortions that are not described by the linear model. Nonlinear models \cite{2012JCAP...11..009V,Okumura:2013zva,Sugiyama:2015dsa} predict the change of sign in the pairwise velocities around separations of about $10 h^{-1}$ Mpc that can be seen in Fig. \ref{fig:coaddDR11}. To show this we implement a nonlinear calculation following \cite{2012JCAP...11..009V,Okumura:2013zva}. The solid line in Fig. \ref{fig:coaddDR11} shows the expected nonlinear model. The evidence for this effect is not particularly strong ($1.8\sigma$ for the 5  Mpc bin) but the low separation points can have an effect on the total S/N. If we fit to the entire range of comoving separations from 0 to 400 Mpc the linear model provides a $3.8\sigma$ detection. Fitting to the nonlinear model recovers the $4.1\sigma$ detection, demonstrating that the low separation points fit better when including nonlinear effects, although the effect on the signal-to-noise is small. These nonlinear models are strongly dependent on the details of the halo model used to calculate them, including the small scale velocity dispersion of dark matter particles. Hence, we do not attempt to extract astrophysical information from this range of separations and we only quote the significance of the measurement for the linear range, but we emphasize that the expected change of sign in the mean pairwise momentum can be clearly seen using spectroscopic data. A proper modeling of these effects will be valuable for future more powerful surveys, and has the potential to probe intra-halo physics. 

The low comoving separation part of the kSZ pairwise estimator might also be affected by the overlap between disks and rings associated with different sources \cite{PhysRevLett.115.191301}. For large separations the overlap between filters can occur as a projection effect; however in this case the overlap between filters simply provides another source of background noise. Moreover, we have found that, for a radius of $1.8\sqrt{2}$ arcmin, the pairs affected by overlapping filters are about $11\%$ of the total in the first bin, $1.5\%$ for the second bin and less than $0.3\%$ for larger separations. Since in this work we discard small separations ($<$ 20 Mpc), the overlap between filters does not have a significant effect on the measured optical depth. This systematic however might need to be modeled or corrected when interpreting data at low separations.

\begin{table*}
\begin{center}
%\hspace*{-1.5cm}
\begin{tabular}{ c |c |c| c | c| c | c}
Luminosity cut/$10^{10}L_{\odot}$ & Mass cut $(M_{200}/10^{13}M_{\odot})$ &N & $\bar{\tau}/10^{-4}$ & S/N Simulations & S/N bootstrap & S/N jackknife \\  
\hline
$L>11.6$ & $M>7.6$  &$ \phantom{a}$5000$\phantom{a}$ & $1.22\pm0.82\phantom{a}$&1.5 & $2.9$ & $2.0$\\ 
$L>9.8$ & $M>6.1$ & $ \phantom{a}$10000$\phantom{a}$ & $1.13\pm0.60\phantom{a}$&1.9 & $4.9$ & $2.1$ \\ 
$L>8.7$ & $M>5.2$    & $ \phantom{a}$15000$\phantom{a}$ &$1.66\pm0.43\phantom{a}$&3.8 & $7.1$ & $3.8$\\ 
$L>7.9$ & $M>4.6$ & $ \phantom{a}$20000$\phantom{a}$ & $1.46\pm0.36\phantom{a}$&4.1 & $8.8$ & $3.7$\\
$L>7.4$ & $M>4.2$ & $ \phantom{a}$25000$\phantom{a}$ & $1.17\pm0.31\phantom{a}$&3.7 & $8.9$ & $3.9$\\
$L>6.9$ & $M>3.8$ & $ \phantom{a}$30000$\phantom{a}$ & $0.99\pm0.28\phantom{a}$&3.6 & $9.0$ & $3.7$\\ 
$L>6.1$ & $M>3.2$ & $ \phantom{a}$40000$\phantom{a}$ & $0.78\pm0.21\phantom{a}$&3.7 & $9.3$ & $3.2$\\ 
$L>5.3$ & $M>2.8$ & $ \phantom{a}$50000$\phantom{a}$ & $0.84\pm0.20\phantom{a}$&3.8 & $10.1$& $4.3$\\
\hline
$7.9<L<9.8$ & $4.6<M<6.1$ & 10000  & $1.42\pm0.68\phantom{a}$&2.1  & $3.6$  & $-$\\
$5.3<L<7.9$ & $2.8<M<4.6$ & 30000  & $0.89\pm0.37\phantom{a}$&2.4   & $2.9$  & $-$\\
%\hline
%$7.7<L<10.5$ & $4.4<M<6.7$ &$\phantom{a}$14253$\phantom{a}$ & $1.64\pm0.48\phantom{a}(3.4)$ & $4.2$  & $-$\\
%$5.3<L<7.7$ &  $2.8<M<4.4$  &$\phantom{a}$28000$\phantom{a}$ & $0.72\pm0.43\phantom{a}(1.6)$  & $2.2$& $-$ \\
\hline
\end{tabular}
\caption{Signal-to-noise and best fit $\bar{\tau}$ as a function of the luminosity cut and of the number of sources (N). The bottom part of the table shows disjoint luminosity ranges. We also report the signal-to-noise ratio for different estimation methods of the covariance matrix. The mass cut is estimated from the halo model and it is used to calculate the mean pairwise velocity curves.}\label{lcut}
\end{center}
\end{table*}

\begin{figure}[h!]
\begin{center}
\includegraphics[width=9cm]{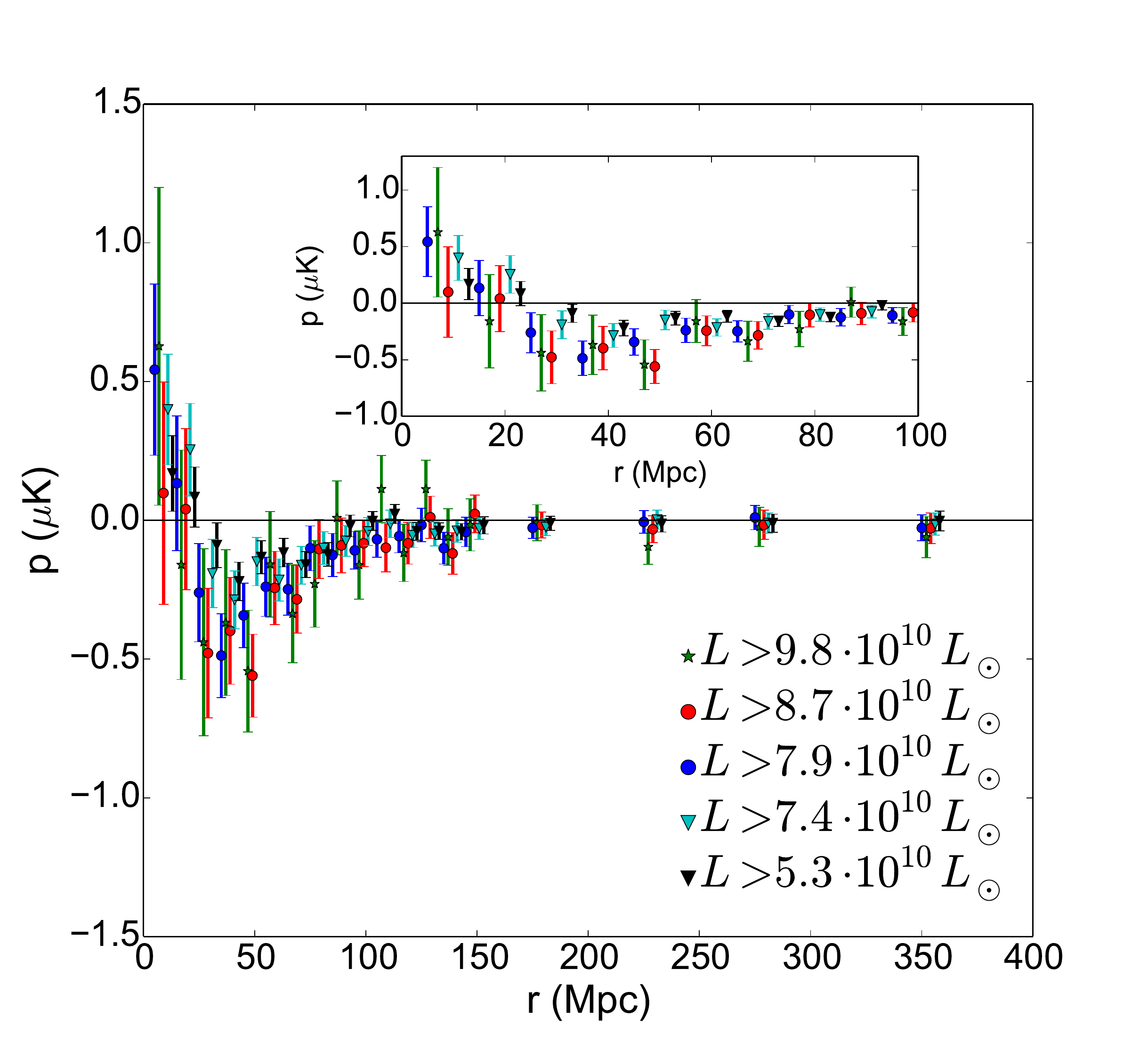}
\caption{Mean pairwise momentum estimator as a function of luminosity cut, see also Table \ref{lcut} for comparison. Each luminosity cut has a horizontal offset for clarity. The inner box shows the same cases in the relevant comoving separation range where the signal is expected. While the negative decrement is always visible, the amplitude of the signal decreases systematically with the minimum luminosity. The signal-to-noise ratio is largest at $L>7.9\times 10^{10}L_{\odot}$ and does not improve significantly for fainter objects (see text for further discussion).}\vspace{-.2in}
\label{fig:all_cuts}
\end{center}
\end{figure}

We have found that the significance of the detection increases consistently with decreasing luminosity cut, down to $L>7.9\times 10^{10}L_{\odot}$. The inclusion of fainter sources in the estimator does not improve the significance of the detection. For example, lowering the luminosity cut to $L>6.9\times 10^{10}L_{\odot}$, corresponding to the $30000$ brightest sources, provides $S/N = 3.6$ and a best fit $\bar{\tau} = (0.99\pm0.28)\times 10^{-4}$. For a luminosity cut of $L>5.3\times 10^{10}L_{\odot}$ (50000 brightest sources) we find $S/N=3.8$ and $\bar{\tau} = (0.84\pm0.20)\times 10^{-4}$. These results are summarized in Table \ref{lcut}. In Fig. \ref{fig:all_cuts} we show the estimator for several luminosity cuts. While the detection remains generally strong, close to $4\sigma$, the overall signal-to-noise ratio does not increase as might naively be expected because of the increased number of pairs. The first 8 bins of Table \ref{lcut} are not independent because all share the most massive sources. To facilitate the comparison between different luminosity ranges we run the analysis on two disjoint samples of the catalog with luminosity $7.9\times 10^{10}<L/L_{\odot}<9.8\times 10^{10}$ and $5.3\times 10^{10}<L/L_{\odot}<7.9\times 10^{10}$. These two cases are listed in the two bottom rows of Table \ref{lcut}. Note that the case reported in the second row of the same table ($L>9.8\times 10^{10}L_{\odot}$) represents another sample that is disjoint from these two ranges. For these cases we have found a significance close to $2\sigma$, specifically $1.9\sigma$ for the most luminous $10000$ sources, $2.1\sigma$ for the next $10000$ sources with $7.9\times 10^{10}<L/L_{\odot}<9.8\times 10^{10}$ and $2.4\sigma$ for the $30000$ sources in the range $5.3\times 10^{10}<L/L_{\odot}<7.9\times 10^{10}$.

%This comparison shows that while the most luminous $\sim10000$ sources do not provide a strong signal ($1.9\sigma$ ) the next $\sim 10000$ sources with $7.9\times 10^{10}<L/L_{\odot}<9.8\times 10^{10}$ provide a significant contribution to the detection (S/N = ) and the fainter sources $5.3\times 10^{10}<L/L_{\odot}<7.7\times 10^{10}$ do not contribute significantly (S/N = )

While the amplitude of the signal is expected to decrease with low luminosity objects, the statistical uncertainty decreases as well, because of the larger number of pairs. The lack of a significant improvement could be caused by systematic effects becoming more dominant for less massive clusters. For example, low luminosity objects are more likely not to lie at the center of clusters and to be satellite galaxies. Mis-centering effects are known to reduce the amplitude of the signal \cite{Flender:2015btu,2016arXiv160303904S} and including a significant amount of satellite galaxies could lead to counting low mass halos multiple times, which would reduce the amplitude of the signal per comoving separation bin. %These effects imply that, while the general behavior of the detetcion significance as a function of the luminosity cut has a physical interpretation, the estimated signal-to-noise can have fluctuations due to noisy sources included in the sample and could be biased with respect to the real signal-to-noise. 

Finally, we note that the best fit optical depth values reported above might include a contribution from gas not bound to the clusters. This has previously been observed when the same filtering approach (aperture photometry) was used for a catalog of central galaxies from the seventh release of the Sloan Digital Sky Survey combined with Planck data \cite{Ade:2015lza,PhysRevLett.115.191301}. In particular Hernandez-Monteagudo et al. (2015) \cite{PhysRevLett.115.191301} find that simulations of the smooth, linear velocity field around the central galaxies provide a better description of the motion of electrons around these galaxies compared to mock catalogs of halos, which describe only the velocity of the halos hosting the central galaxies. The optical depth estimates in Table \ref{lcut} are consistent with those reported in \cite{PhysRevLett.115.191301}, which, for similar physical apertures and similar halo masses, estimated $\tau$ in the range $0.2-2 \times10^{-4}$. This is encouraging and might imply that the pairwise kSZ signal detected in this work is also sensitive to the motion of electrons not bound to the clusters. We note however that a detailed direct comparison between the $\tau$ values of our paper and those of \cite{PhysRevLett.115.191301} is complicated by the different redshift population, which is limited to $z\sim 0.1$ for the sample used in \cite{PhysRevLett.115.191301}, and by the different angular resolutions involved.

\subsection{redMaPPer}\label{sec:redmapper}
redMaPPer is a red-sequence cluster finder algorithm, which provides cluster center positions based on the best esimates for the position of the central galaxy. As described above, the redMaPPer SDSS DR8 catalog has about 31683 clusters in the ACT+ACTPol area. To account for incompleteness and masking effects, redMaPPer recommends using a corrected richness, $\lambda_s = \lambda/s$, where $s$ is a correction factor. In our region we find only $2242$ clusters with richness $\lambda_s>20$ and with an average redshift of 0.35. These are shown in Fig. \ref{fig:catalogs}. Clusters with $\lambda_s<20$ have larger uncertainties on the richness and we do not include them in the analysis. Given the reduced number of clusters above the $\lambda_s=20$ threshold we did not find a significant detection of the kSZ signal using this catalog. This stresses the importance of having deep optical catalogs to overlap with the CMB maps for these kinds of studies. 

\subsection{Null tests}
We performed several null tests to confirm that the signal is not due to systematic effects in the maps or in the analysis. The simplest null test can be performed by transforming the difference in (\ref{estimator_t}) into a sum so that the kSZ will average to zero like all the other contributions to the temperature fluctuations associated with the clusters. The fit to the template provides an amplitude consistent with the null signal as expected, $\tau_{\rm null}= (-0.07\pm3.5)\times 10^{-5}$, and a $\chi^2$ of 13 for 19 degrees of freedom (dof) with a probability to exceed (PTE) of 0.84.

The CMB simulated maps that we use for the covariance matrix provide a more stringent null test. The average of 400 null tests obtained by applying the estimator to these mock maps of CMB and noise provides  $\tau_{\rm null} =(-0.14\pm0.19)\times 10^{-5}$ and a $\chi^2$ of 16 for 19 dof (PTE = 0.66), again consistent with no signal. 

\begin{figure}[!t]
\begin{center}
\includegraphics[width=9cm]{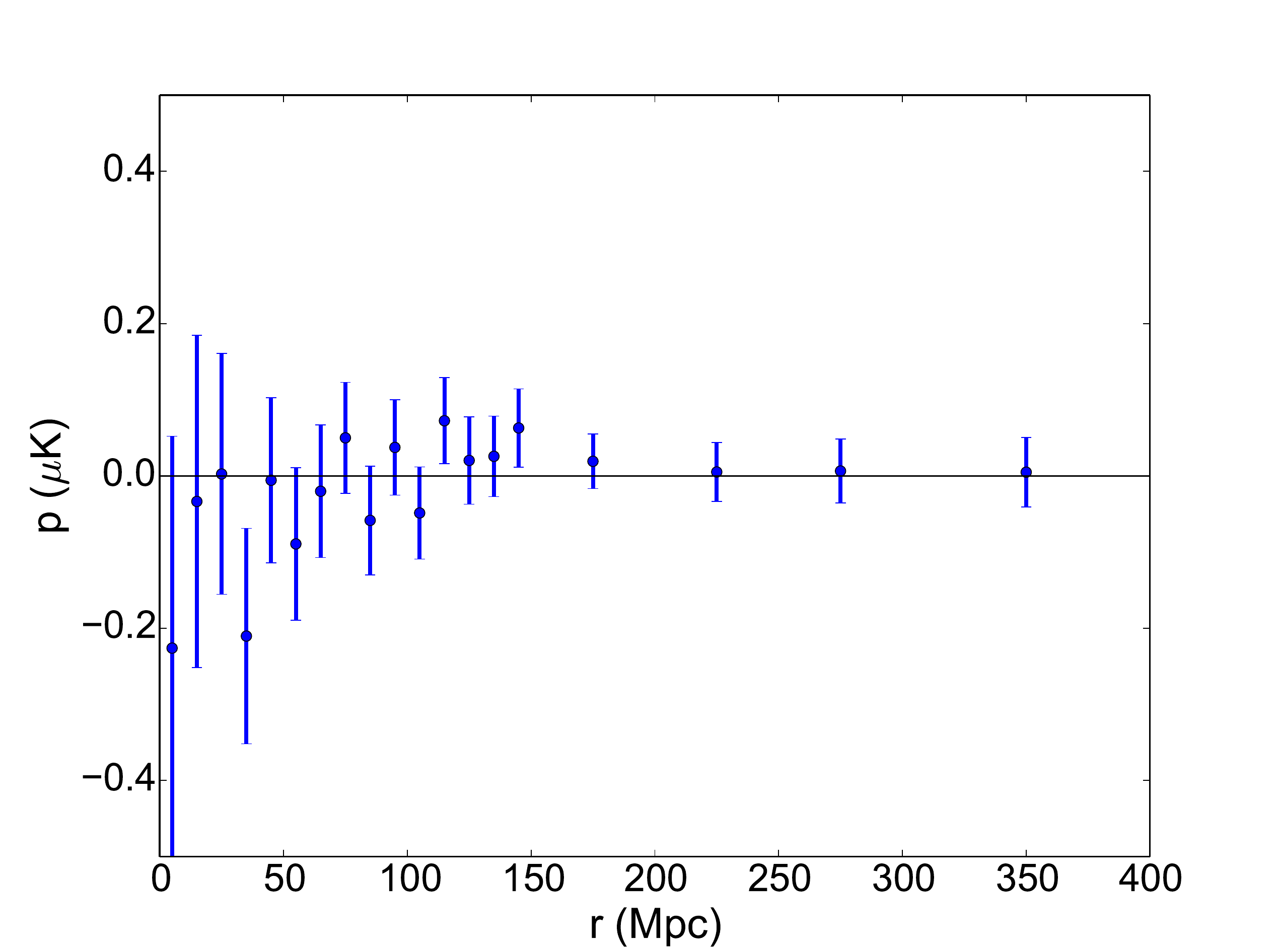}
\includegraphics[width=9cm]{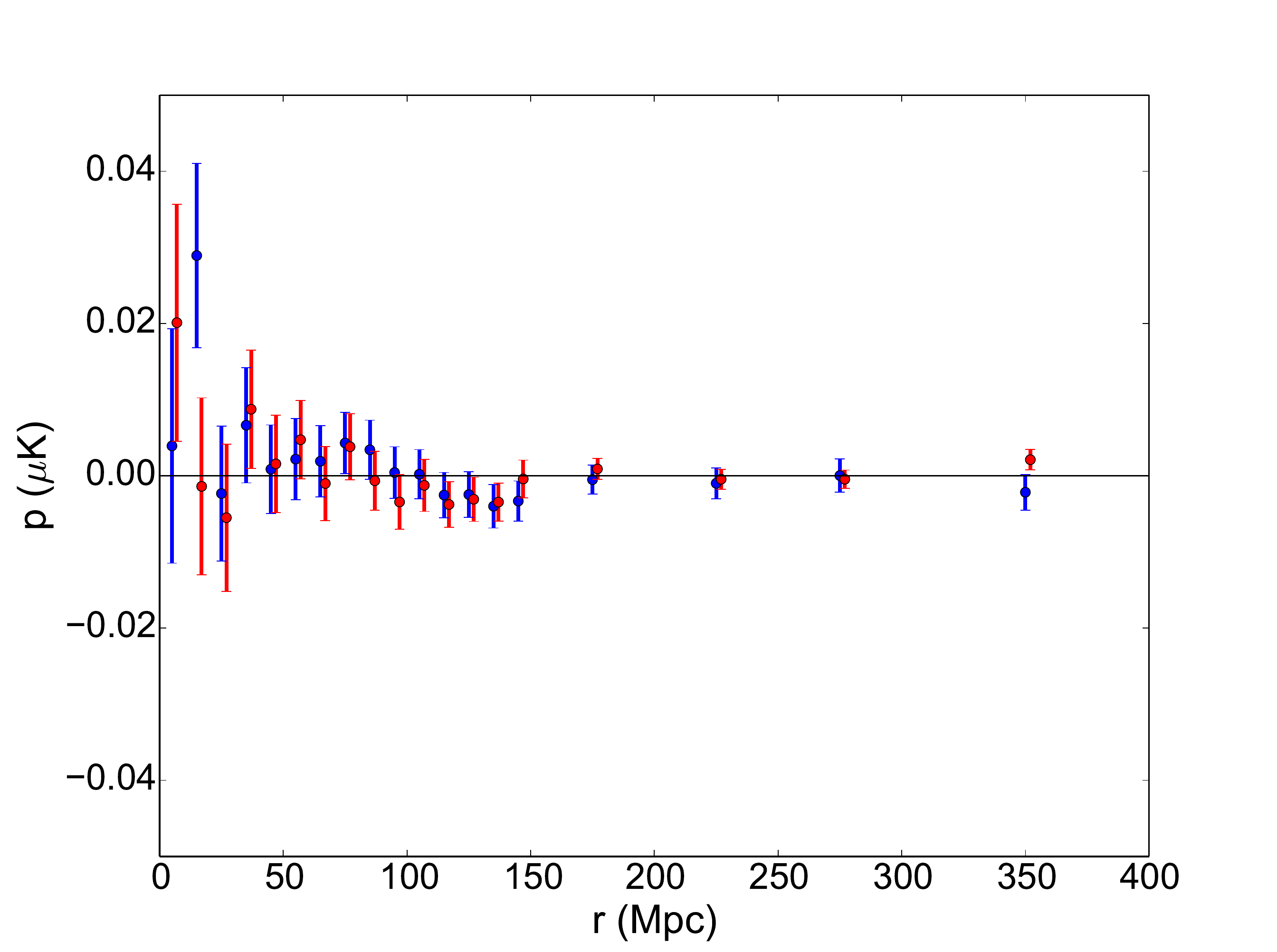}
\caption{Null tests. Top: Same as in Fig. \ref{fig:coaddDR11} but changing the sign in (\ref{estimator_t}). In this case we find $\tau_{\rm null}= (-0.07\pm3.5)\times 10^{-5}$. Bottom: Average of 400 null tests obtained from mock maps of CMB and noise (blue). We find $\tau_{\rm null} =(0.10\pm0.11)\times 10^{-5}$. The red points are the average of 400 null tests obtained by shuffling the temperature values, keeping the sources fixed at their positions. In this case we find: $\tau_{\rm null}= (-0.05\pm0.14)\times 10^{-5}$. All these null tests are consistent with zero signal as expected.}\vspace{-.2in}
\label{fig:null}
\end{center}
\end{figure}

Finally, we randomly shuffle the temperature values of the sources, keeping their position fixed (and hence the $c_{ij}$ weights of the estimator), and take the average of these realizations. This case is also consistent with no signal: $\tau_{\rm null}= (0.06\pm0.14)\times 10^{-5}$ and $\chi^2=17$ for 19 dof (PTE = 0.59).
The null tests are shown in Fig. \ref{fig:null}.
 
\begin{figure}[!t]
\begin{center}
\includegraphics[width=9cm]{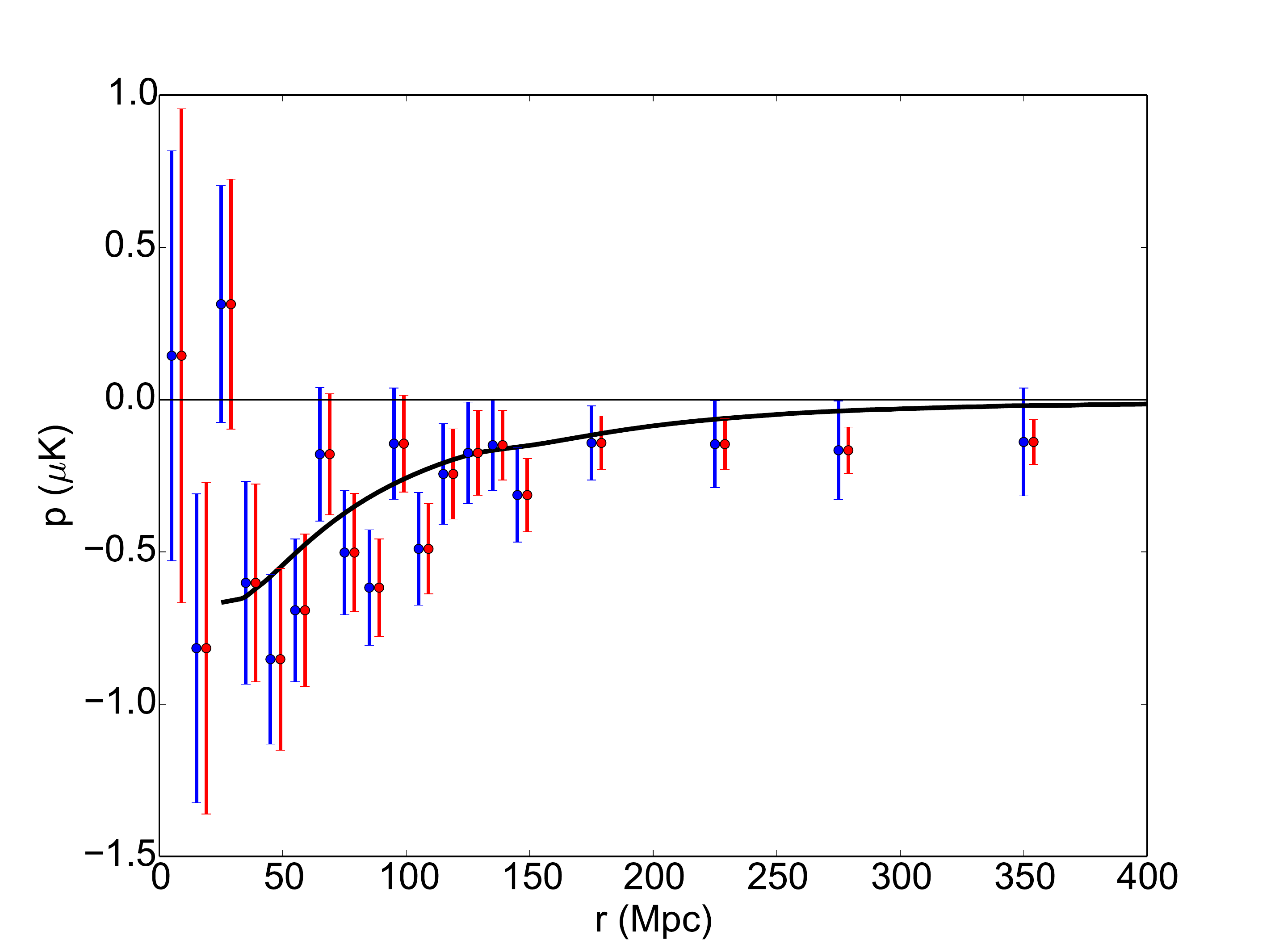}
\includegraphics[width=9cm]{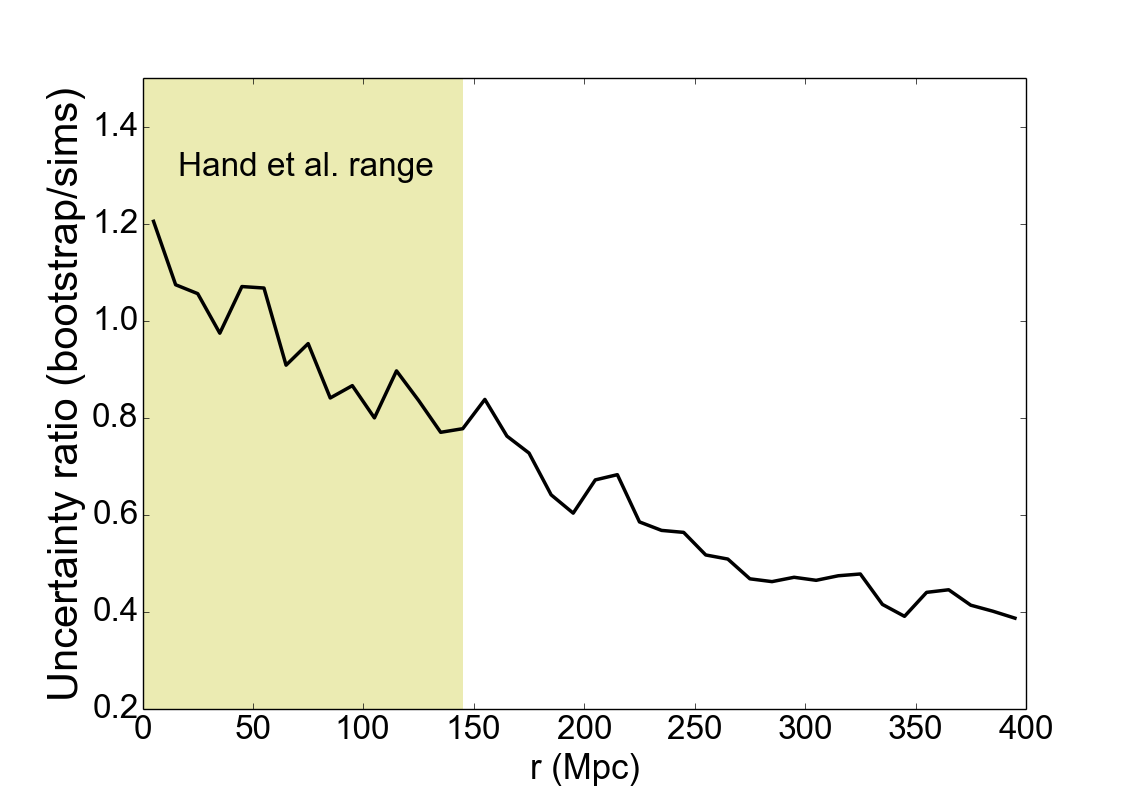}
\caption{\textit{Top}: Mean pairwise momentum from the $7800$ most luminous ($L>7\times 10^{10}L_{\odot}$) DR9 galaxies in the ACT region for a $1.7$ arcmin aperture photometry filter. We used 40 comoving separation bins for the analysis, but the large separations points are strongly correlated and we re-binned them for plotting purposes. We show error bars from mock CMB maps (blue) and bootstrap (red). Note that the bootstrap uncertainty estimates are much smaller at large separations. \textit{Bottom}: Ratio between error bars from the bootstrap approach to those from simulations. The error bar differences range between $10\%$ and $20\%$ up to $150$ Mpc, which was the maximum separation used by H12 in \cite{2012PhRvL.109d1101H}.}\vspace{-.2in}
\label{fig:ACT_HandCat}
\end{center}
\end{figure}

\begin{figure}[!t]
\begin{center}
\includegraphics[width=9cm]{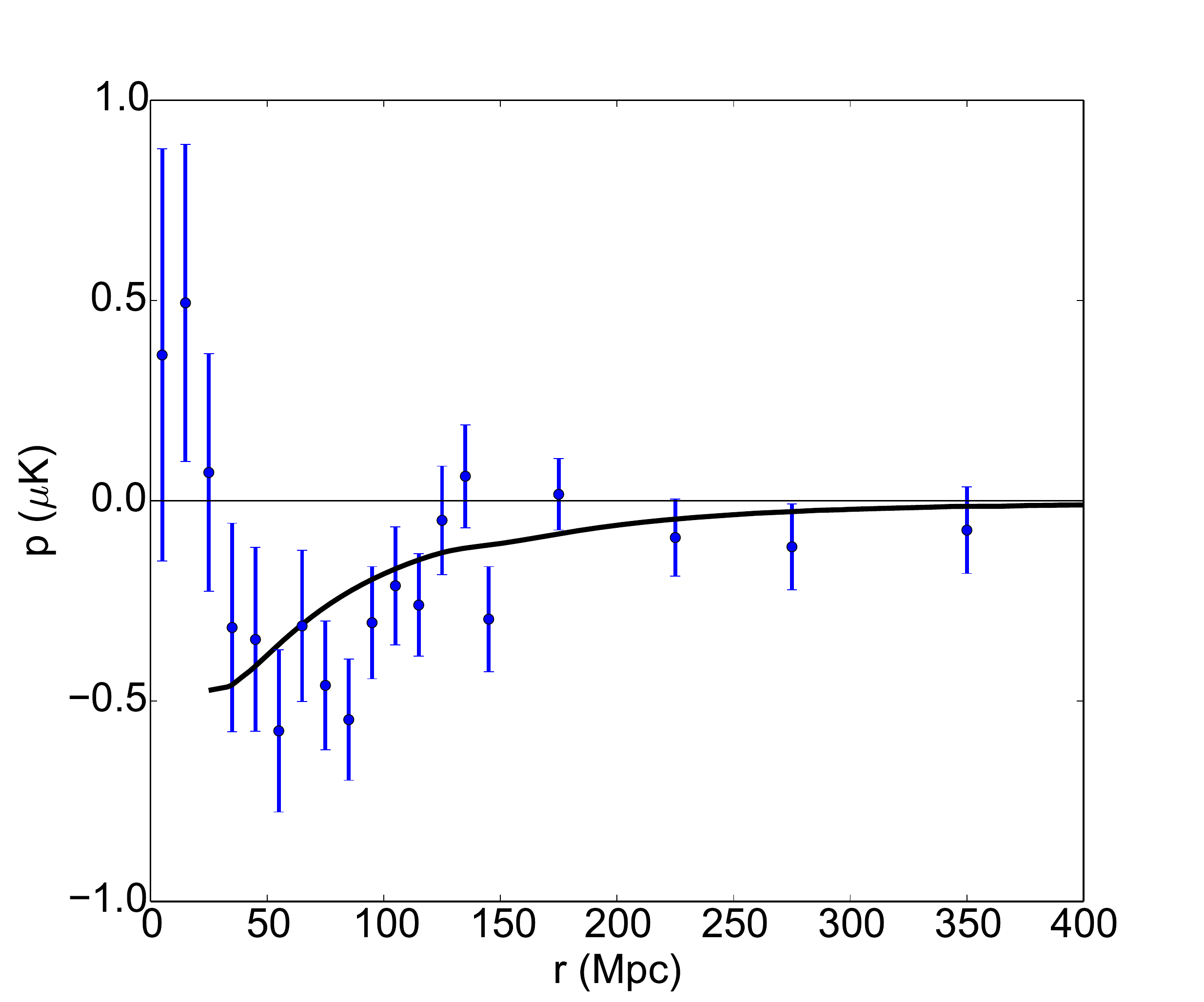}
\caption{Mean pairwise momentum from the $9800$ most luminous ($L>7.3\times 10^{10}L_{\odot}$) DR11 galaxies in the ACT region for a $1.7$ arcmin aperture photometry filter. Error bars are estimated from mock CMB maps.}\vspace{-.2in}
\label{fig:ACT_DR11}
\end{center}
\end{figure}

\section{Comparison with the  previous ACT analysis}\label{sec:comparison}
This analysis is an extension of the work presented by H12 in \cite{2012PhRvL.109d1101H}. In that work, the authors used a 220 deg$^2$ map from ACT, combined with a selection of the BOSS DR9 galaxy catalog, and reported the first evidence for the kSZ pairwise momentum, rejecting the null signal with a significance of 3.1$\sigma$.

Despite the 2.7 times wider area used in this paper and the lower map noise, the evidence for the pairwise momentum presented in the previous section is not substantially higher than the one found by H12. This is due to a combination of the more conservative covariance matrix calculation and map filtering approaches and the different optical catalog used. 
In H12 the CMB maps were filtered with a matched filter (MF) assuming that the cluster profile was described by a Gaussian with a $1.4$ arcmin FWHM (similar to the ACT beam profile) and the covariance matrix was calculated with a bootstrap resampling. Here we adopt the more conservative model independent aperture photometry (AP) filter described in \ref{sec:filt} and estimate the covariance matrix using simulations. 

To compare with H12, we apply our approach to the same ACT region used in H12, covering 220 deg$^2$ with right ascension ranging from $-43^{\circ}$ to $+45^{\circ}$ and declination $-1.25^{\circ}$ to $+1.25^{\circ}$ and the same selection of DR9 sources. H12 reported evidence of the kSZ pairwise momentum in the range of separations 0-150 Mpc, with a probability of the signal being due to random errors of $p=0.002$, corresponding to a $3.1\sigma$ rejection of the null signal. After exploring various luminosity cuts we have confirmed this result using the AP filter and estimating the covariance matrix using simulations. For the same range of comoving separations used by H12(0-150 Mpc) the null signal is rejected at $3\sigma$, consistent with the result reported by H12. 

In Fig. \ref{fig:ACT_HandCat} we show the results of this new analysis with 7800 objects (instead of the 5000 used in H12), corresponding to a minimum luminosity of $L > 7\times 10^{10} L_{\odot}$. The filter aperture used for this analysis is $1.7$ arcmin, consistent with the typical cluster size in the sample considered, and the significance is stable against small variations of this aperture. By fitting to the pairwise kSZ  velocity template we find an amplitude $\bar{\tau} = (3.8\pm 0.9)\times 10^{-4}$. The covariance matrix calculation is based on simulations and it is one of the main differences between this work and the H12 paper, where the bootstrap approach was used to estimate the uncertainties. Fig. \ref{fig:ACT_HandCat} shows the results for this analysis with error bars estimated with the two methods. As described above, the uncertainties from bootstrap are between 10$\%$ and 20$\%$ smaller than those calculated with simulations for separations $r<150$ Mpc. The difference becomes more relevant at larger separations. 

%With the bootstrap method we find a significance close to 9$\sigma$ when fitting to the range $r = 20-400$ Mpc, although with a reduced $\chi^2_{red}=2.5$ for the best fit model, which suggests underestimated uncertainty.

To study the effect of changing the optical sample, we have repeated the same analysis using the same ACT map used by H12 combined with the DR11 LSS catalog (Fig. \ref{fig:ACT_DR11}). As described above this catalog has about 1000 fewer objects and more low redshift sources compared to the DR9 H12 selection. We find that the maximum signal to noise ratio is achieved for a slightly different luminosity cut, $L>7.3\times 10^{10}L_{\odot}$, corresponding to the $9800$ brightest objects in the catalog, for the same filter aperture of $1.7$ arcmin. The best fit amplitude for the model template is $\bar{\tau} = (2.7\pm 0.7)\times 10^{-4}$,
%corresponding to a $3.6\sigma$ detection,
slightly less significant than the same analysis using the DR9 catalog. The rejection of the null signal is also lower: $2.6\sigma$. Since both the CMB map and the filter scheme are the same we conclude that the different significance achieved is related to the differences between catalogs. The luminosity cut for the DR9 analysis provides sources with an average redshift of $z=0.56$, while for the DR11 cut we find $z=0.48$. The higher number of low-redshift sources in DR11 might imply that a larger fraction of these sources are satellite galaxies that do not properly trace the center of clusters. This difference stresses the dependence on the catalog and the importance of the optical sample used for the kSZ pairwise analysis.

\section{Optical depth from the thermal SZ signal}\label{sec:tsz}
\subsection{Overview}
Reconstructing pairwise velocities from the mean pairwise momentum estimator requires knowledge of the optical depth $\tau$ of the objects used for the statistics. In the kSZ effect this is completely degenerate with the peculiar velocity itself. Hence, estimating peculiar velocities from the kSZ requires including additional information. The thermal component of the SZ effect (tSZ) has a dependence on the optical depth and the electron temperature $T_e$, making it possible to measure the tSZ effect for the same sources used for pairwise statistics and infer $\tau$ from the tSZ signal \cite{Sehgal:2005cy}. This measurement can then be used to convert the pairwise momentum into a pairwise velocity. This section shows an example of this approach.

A direct measurement of $\tau$ from the tSZ effect would require either assuming a temperature or estimating $T_e$, to break the $\tau$-$T_e$ degeneracy. Here we use a different approach, by measuring the average tSZ signal from the same sample used for the kSZ analysis by stacking on the positions of sources belonging to the same luminosity bins. We apply a matched filter to the entire coadded ACTPol-ACT map using a cluster profile template. The matched filter provides an estimate of the central SZ temperature decrement associated with a cluster. We use this central value to normalize the cluster profile, which, by assumption, has the same shape as the one used for the matched filter. We then estimate the comptonization parameter in a 1.8 arcmin circle, that is, the same aperture used for the kSZ analysis above and use a theoretical relation between optical depth and Comptonization parameter from hydrodynamical simulations of clusters to infer the optical depth (Battaglia (2016) \cite{2016arXiv160702442B}). Finally, we compare this tSZ estimated $\tau$ with the one measured from the kSZ pairwise momentum. The next section discusses the limits and possible systematic effects associated with this approach.

%The recovered temperature decrement is converted into a comptonization parameter $y$, and, by using results from hydrodynamical simulations (Battaglia (2016) \cite{2016arXiv160702442B}), into a $\tau$ estimate. We then compare this tSZ $\tau$ with the $\bar{\tau}$ from pairwise kSZ measurements and discuss the differences.
\subsection{Thermal SZ}
For tSZ analysis the best signal-to-noise is obtained by filtering the CMB map with a matched filter. We filter our coadded map using the same matched filter approach used in Hasselfield et al. \cite{Hasselfield:2013wf}, based on a Universal Pressure Profile \cite{2010A&A...517A..92A}, with a fixed scale of $\theta_{500}=5.9$ arcmin (see also discussion below). After the filter is applied, the sources from the DR11 catalog are binned using the same luminosity bins as Table \ref{lcut}. For each luminosity bin, a 10.5 arcmin by 10.5 arcmin submap centered on each source is repixelized from 0.5 arcmin per pixel to 3.75 arcsec per pixel. Resizing the pixels minimizes the errors associated with the relatively large pixel size of the ACTPol-ACT maps (0.5 arcmin) \cite{2011ApJ...736...39H}. The temperature decrement associated with each source is taken to be the central pixel value in $\mu$K. These decrements are averaged within each luminosity bin to obtain a stacked tSZ signal per bin, $\delta T_{\rm tSZ}$. The error associated with each temperature decrement is obtained by taking the standard deviation of the pixels within an annulus of inner radius $R_{1} = 3$ arcmin and an outer radius of $\sqrt2R_{1}$. The 3 arcmin size of this annulus is a conservative estimate of the local noise in the map. This approach has the advantage of not requiring the modeling of variations of the noise across the map. We have verified that the distribution of the standard deviation values is similar for annuli around the sources and for annuli selected in source free regions, implying that, regardless of the presence of sources, there can be non-negligible variations in the noise across the map. We have also verified that the error increases monotonically up to 3 arcmin and is stable for larger rings. In Table \ref{table_tSZ} we present the $\delta T_{\rm tSZ}$ per luminosity bin.\\

To obtain the comptonization parameter we follow the steps detailed in \cite{Hasselfield:2013wf}. The tSZ temperature signal is related to the Comptonization parameter $y$ by:

\begin{equation}\label{y0}
\frac{\Delta T(\theta)}{T_{\rm CMB}} = f_{\rm SZ}\hspace{3pt}y(\theta)  ,
\end{equation}
where $y(\theta)$ is the Compton parameter at a projected angle $\theta$ from the cluster center, and in the non-relativistic limit, $f_{\rm SZ}$ depends on observed radiation frequency alone. At an effective frequency of 146.9 GHz, $f_{\rm SZ}=-0.992f_{\rm rel}(t)$ where $f_{\rm rel}(t) = 1 + 3.79x - 28.2x^2$ and $x = k_BT_e/m_ec^2$. A cluster gas temperature of $T = 0.5$ keV is assumed, and $f_{\rm SZ}$ varies minimally with this choice of temperature. For each source, $y(\theta = 0)$ is obtained from the central pixel temperature decrement via equation (\ref{y0}), and the averaged $\bar{y}$ per luminosity bin is reported in Table \ref{table_tSZ}. \\

The angular averaged Compton parameter $\bar{y}$ is found for each source by integrating over the generalized Navarro-Frenk-White (GNFW) pressure profile \cite{Nagai:2007mt},

\begin{equation}
y(\theta) \propto\int ds P\Big(\sqrt{s^2 + (R_{500} \theta/\theta_{500})^2}\Big),
\end{equation}
where $\theta_{500} = R_{500}/D_A(z)$, $D_A(z)$ is the angular diameter distance to the source with redshift $z$ and we vary $R_{500}$ to fix $\theta_{500}=5.9$ arcmin. The integral $s$ is along the line of sight, with $P(r)$ being the pressure profile, defined as in \cite{Hasselfield:2013wf}. We normalize this integral with the $y_0$ value of each luminosity bin and calculate an averaged Compton $\bar{y}_{\theta}$ parameter:

\begin{equation}\label{compint}
\bar{y}_{\theta} = \frac{2}{\theta^2}\int_0^{\theta} y(\theta^{\prime})\theta^{\prime} d\theta^{\prime}. 
\end{equation}

A recent analysis of cluster hydrodynamical simulations has found a strong relationship between the averaged $\bar{y}$ and the cluster optical depth (Battaglia (2016), \cite{2016arXiv160702442B}).  Specifically, in simulations with AGN feedback Battaglia (2016) finds $\ln(\tau) = \ln(\tau_0) + m\ln(\bar{y}/10^{-5})$, where $\ln(\tau_0) =-6.40$ and m = $0.49$ at z = 0.5.

We check the viability of using the $\tau$ inferred from tSZ measurements and hydrodynamical simulations to obtain an estimate of the mean pairwise velocity from the mean pairwise momentum. Unless systematic biases are present in the analysis, the optical depth obtained by fitting to the pairwise velocity template should be consistent with the one estimated from the thermal SZ within the same angular aperture. To quantify the agreement between these measurements we fit a linear relation to the three independent $\tau$ estimates, corresponding to the first bin and the last two bins in Table \ref{lcut}, accounting for the uncertainties both on the kSZ and the tSZ measurements. Battaglia (2016) provides fitting relations for the angular averaged comptonization parameter (\ref{compint}) and the optical depth averaged in circular areas of radii 1.3, 1.8, 2.6 and 5.2 arcmin. These hydrodynamical simulations also provide an estimate of the systematic uncertainty on the best fit parameters of the $\bar{y}$-$\tau$ relation, accounting for differences between radiative cooling and AGN feedback models. For $1.8$ arcmin, the same aperture used for the kSZ analysis of this paper, the uncertainties are $4\%$ and $8\%$ respectively on $\ln{\tau_0}$ and $m$. %We combine these uncertainties in quadrature with the statistical uncertainty on $\tau$ estimated from the tSZ.  

\begin{figure}[h!]
\begin{center}
\includegraphics[width=9cm]{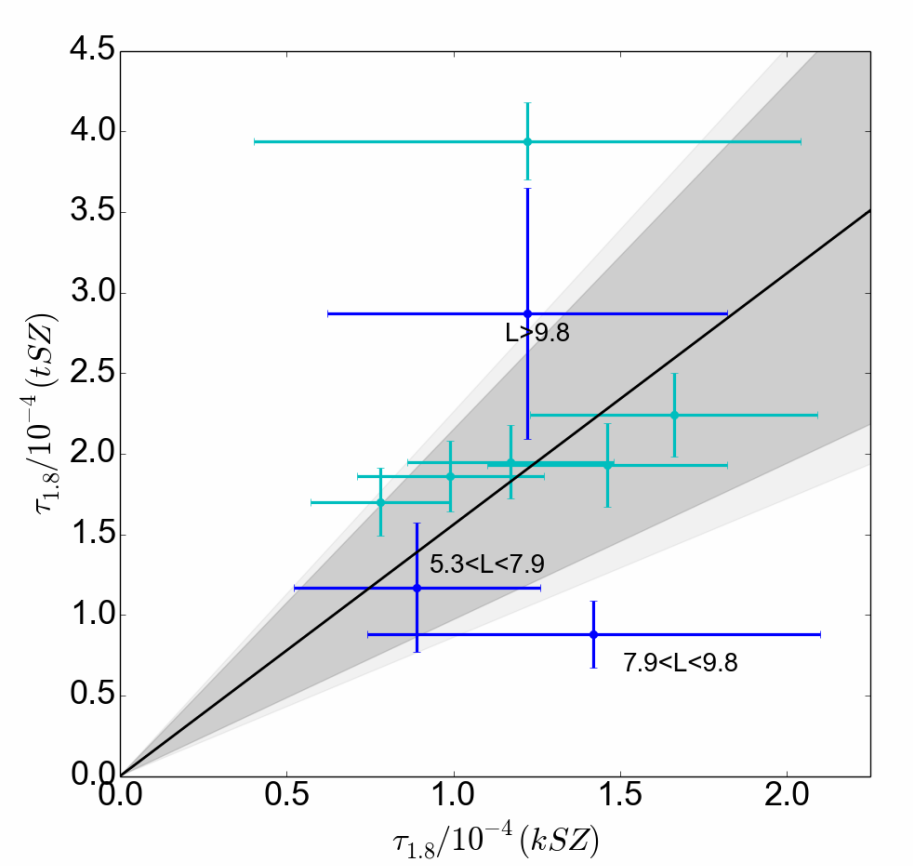}
\caption{Best-fit line for the tSZ and kSZ optical depth measurements. We only fit to the dark blue points, which correspond to the three independent luminosity bins (second row and and last two rows in Table \ref{lcut}) indicated by the labels (in units of $10^{10}L_{\odot}$). The cyan points correspond to the remaining bins. Fitting to the 1.8 arcmin $\bar{y}$-$\tau$ relation from hydrodynamical simulations we find a slope $m_{1.8}=1.60 \pm 0.49(\rm stat)\pm 0.59(\rm sys)$. The gray areas represent the $1\sigma$ statistical (dark) and systematic (light) uncertainty range.}
\label{fig:mfits}
\end{center}
\end{figure}

We compare the kSZ optical depths to those obtained from tSZ measurements using the $\bar{y}$-$\tau$ relation and find a slope $m_{1.8}=1.60 \pm 0.49(\rm stat)\pm 0.59(\rm sys)$  (see Fig. \ref{fig:mfits}) where the systematic uncertainty is related to the uncertainty on the best fit parameters for the $\bar{y}$-$\tau$ relation. The resulting slope is within $1.3\sigma$  and $1\sigma$ of unity considering the statistical and systematic uncertainties respectively, suggesting that the tSZ and kSZ optical depth estimates are consistent given current uncertainties and that this approach is promising for extracting pairwise velocities from pairwise momentum measurements. Improved data are needed to perform a more detailed search for systematics, such as mis-centering, dusty galaxy contamination of the tSZ signal, modeling of the pairwise velocities, or filtering of the maps. We note that a potential source of bias is that the hydrodynamical simulations provide the $\bar{y}$-$\tau$ relationship for clusters with masses $M_{500c} = 9\times 10^{13} M_{\odot}$ at $z=0.5$, higher than the mass range predicted from the halo model for the clusters used in this analysis. 

In Table \ref{table_tSZ} we summarize the tSZ temperature decrement, the averaged comptonization parameters for an aperture of 1.8 arcmin and the corresponding optical depth, as well as the uncertainty on the optical depth from simulations for each luminosity bin. 
We rescale the best fit pairwise momentum of Fig. \ref{fig:coaddDR11} using the tSZ estimated optical depth for the same luminosity cut ($\tau_{tSZ,1.8}=(1.93\pm0.26)\times 10^{-4}$) and find a mean pairwise velocity of $145\pm 40(\rm stat) \pm 72(\rm sys)$ km/s at separation 35 Mpc. This is within $1\sigma$ of the $\Lambda$CDM model prediction for the same mass and average redshift of the sample considered, $v_{\Lambda CDM} = 193$ km/s. 

\begin{table*}
\begin{center}

\begin{tabular}{c|c|c|c|c|c|c|c|c}
Luminosity cut/$10^{10} L_{\odot}$ & N & $<L>$ & $<$z$>$ & $\delta T_{tSZ}$ ($\mu$K) & $y_0/10^{-7}$ & $\bar{y}_{1.8}/10^{-7}$ & $\tau_{1.8}/10^{-4}$  & $\sigma_{sims}/10^{-4}$\\
%\cline{3-6}
\hline
$L > 11.6$      &\quad 4650  \quad& 14.90 &\quad 0.52 \quad&\quad -3.70 $\pm$ 0.49 \quad       &\quad 13.60 $\pm$ 1.82 \quad         & 6.09$\pm$0.81            & 4.22$\pm$0.28             &$\pm$1.11\\
$L > 9.8$        &\quad 9269  \quad& 12.73 &\quad 0.51 \quad&\quad -1.67  $\pm$ 0.35 \quad       &\quad 6.17  $\pm$ 1.29 \quad         & 2.78$\pm$0.58           & 2.87$\pm$0.29             &$\pm$0.78\\
$L > 8.7$       &\quad 13898 \quad& 11.56 &\quad 0.50 \quad&\quad -1.23 $\pm$ 0.29 \quad       &\quad 4.52 $\pm$ 1.06 \quad          & 2.05$\pm$0.48           & 2.47$\pm$0.28             &$\pm$0.68\\
$L > 7.9$       &\quad 18586 \quad& 10.75 &\quad 0.49 \quad&\quad -0.91 $\pm$ 0.25 \quad       &\quad 3.35 $\pm$ 0.92 \quad          & 1.52$\pm$0.42           & 2.14$\pm$0.29             &$\pm$0.60\\
$L > 7.4$       &\quad 23251 \quad& 10.13 &\quad 0.48 \quad&\quad -0.90 $\pm$ 0.22 \quad       &\quad 3.32 $\pm$ 0.82 \quad          & 1.51$\pm$0.37           & 2.13$\pm$0.26             &$\pm$0.60\\
$L > 6.9$       &\quad 27877 \quad& 9.64 &\quad 0.48 \quad&\quad -0.79  $\pm$ 0.20 \quad        &\quad 2.92  $\pm$ 0.75 \quad         & 1.34$\pm$0.34           & 2.01$\pm$0.25             &$\pm$0.58\\
$L > 6.1$       &\quad  37190 \quad& 8.85 &\quad 0.47 \quad&\quad -0.65 $\pm$ 0.18 \quad         &\quad 2.41 $\pm$ 0.65 \quad         & 1.11$\pm$0.30            & 1.83$\pm$0.24             &$\pm$0.54\\
$L > 5.3$       &\quad  46448 \quad& 8.22  &\quad 0.47 \quad&\quad -0.52 $\pm$ 0.16 \quad          &\quad 1.91 $\pm$ 0.58 \quad       & 0.88$\pm$0.27         & 1.64$\pm$0.24             &$\pm$0.49\\
\hline
$7.9 < L < 9.8$ &\quad 9299 \quad& 8.77 &\quad 0.45 \quad&\quad -0.15 $\pm$ 0.35 \quad    &\quad 0.54 $\pm$ 1.30 \quad         & 0.25$\pm$0.60           & 0.88$\pm$1.04              &$\pm$0.21\\
$5.3 < L < 7.9$ &\quad 27880 \quad& 6.53 &\quad 0.45 \quad&\quad -0.26 $\pm$ 0.20 \quad       &\quad 1.00 $\pm$ 0.75 \quad        & 0.45$\pm$0.35          & 1.17$\pm$0.45               &$\pm$0.40\\
\hline
\end{tabular}
\caption{Extracted tSZ temperature decrements $\delta T_{\rm tSZ}$, central Compton parameter $y_0$, angular averaged $\bar{y}$ for a 1.8 arcmin radius circle and estimated optical depth $\tau$ for the same luminosity ranges as Table \ref{lcut}. The number of sources (N) considered per luminosity cut is listed along with average luminosity $\langle L\rangle$ and average redshift $\langle$z$\rangle$ for the sample. Note that the number of sources per luminosity bin differs from those in Table \ref{lcut} because the wider submaps needed for tSZ stacking require discarding more sources at the edges of the map.}\label{table_tSZ}
\end{center}
\end{table*}

%\begin{figure}[h!]
%\begin{center}
%\includegraphics[width=9cm]{../plots/pairwiseV.png}
%\caption{Mean pairwise velocity obtained from the mean pairwise momentum measured in Figure \ref{fig:coaddDR11} and assuming an optical depth estimated from the tSZ stacking analysis for the same luminosity cut. The shaded areas represent the 1$\sigma$ (dark gray) and 2$\sigma$ (light gray) ranges.}\label{pairwiseV}
%\end{center}
%\end{figure}

\section{Systematic Effects}\label{sec:sys}
Several systematic effects can affect the amplitude of the kSZ signal. One of the main potential sources of error when using galaxies to trace clusters is the mis-centering of the galaxies with respect to the cluster centers. The details of the sub-halo structure can affect the measured pairwise momentum especially at low separations and is not easily modeled. The filtering of the CMB map also can affect the recovered S/N. Flender et al.\cite{Flender:2015btu} have quantified several of these effects using N-body simulations and found that the mis-centering of sources can reduce the amplitude of the signal at most by $10\%$ at all separations, in a pessimistic scenario. They have also found that star formation and feedback can reduce the amplitude of the pairwise kSZ by $\sim50\%$. Comparing the AP filter with a matched filter they have found that the former could be slightly more conservative than the latter, providing a S/N from 10$\%$ to $18\%$ lower, depending on the kSZ model. 

Imperfect removal of other effects, especially of the tSZ, can bias the pairwise kSZ signal. For the pairwise statistic this could happen if there are not enough pairs to achieve a perfect cancellation of the tSZ effect, especially for the most massive clusters. Schaan et al.\cite{Schaan:2015uaa} found that, for the reconstructed velocity method, removing the most massive clusters ($M>10^{14}M_{\odot}$) is important to avoid contamination. The sample used in this paper has mostly clusters with masses below that range. Based on halo model estimated masses we have found only 237 objects with $M>10^{14}M_{\odot}$. Removing those objects from the sample does not affect the significance of the pairwise kSZ measurement.

For the thermal SZ stacking, a potential systematic is emission from dusty star-forming galaxies. At the 148 GHz frequency used for this analysis the dust emission is not negligible \cite{Greco:2014vwa,2013A&A...557A..52P}. Data at 220 GHz, where the tSZ effect vanishes, can be used to estimate the dust contribution and remove it from the measured temperature decrement at 148 GHz with an appropriate frequency scaling (see for example \cite{2010ApJ...718..632H}). We do not yet have measurements at 220 GHz covering the full coadded ACTPol + ACT map. A 220 GHz map is available for the noisier ACT only region. Using these data we have found that removing the dust can increase the value of the estimated tSZ optical depth by about $20\%$ for luminosities $>7.9\times10^{10}L_{\odot}$ and up to $50\%$ for the lower luminosity bin, where the tSZ signal is weaker. Estimating the dust contribution will hence be very important for future data. The ongoing Advanced ACTPol \cite{Henderson:2015nzj} survey will soon provide multi-frequency maps covering a wider region, allowing for an appropriate analysis of the dust contamination. 

Another potential source of systematics is the $\bar{y}$-$\tau$ conversion.  In addition to the quoted systematic uncertainty between different hydrodynamical simulations, there is also uncertainty in extrapolating from the larger masses in the hydrodynamical simulations to the lower mass objects in our sample. Future simulations extending to lower masses can address this uncertainty. Alternatively, wider surveys with deeper optical data will allow the use of larger catalogs, providing a significant number of sources with masses in the same range as the current simulations. We also observe that the $\bar{y}$-$\tau$ relation provided by Battaglia (2016) does not involve any filtering of the simulated clusters. To assess whether the matched filter affects the measured $\tau$ we apply the same matched filter used in our analysis to the projected simulated maps used by Battaglia (2016), without instrumental noise nor CMB backround. We find that the mean optical depth estimated after applying the matched filter is about $30\%$ larger than the optical depth without filtering. A detailed investigation of this bias will require more extensive work. However, under the pessimistic assumption that all the optical depths estimated with the matched filter are biased $30\%$ high, we find that the slope for the tSZ-kSZ optical depths fit is in better agreement with unity: $m = 1.20\pm 0.32(\rm stat)\pm0.51(\rm sys)$. This value is well within the error bars of the best fit quoted in the previous section. We conclude that, at this stage, the approach used in this paper is dominated by statistical and systematic uncertainties. Addressing proper filtering of the simulated maps and how the $\bar{y}$-$\tau$ relation is affected by the filter should be addressed before applying this approach to data from future surveys. 
 
In this analysis we used the best fit relation provided by Battaglia (2016) for a redshift $z=0.5$. This is a reasonable choice, because, as shown in Table \ref{table_tSZ}, the average redshift of all the luminosity bins is close to $0.5$. We have checked the redshift dependence of the $\bar{y}$-$\tau$ conversion by repeating the analysis for the best fit values provided by Battaglia (2016) for $z=0.3$ and $z=0.7$. We have found variations in the value of $\tau$ in the range $6$-$20\%$. These variations are still within $1$-$2\sigma$ of our estimates but suggest that future larger surveys, which will be able to split the cluster sample into redshift bins, will need to account for the redshift dependence of the $\bar{y}$-$\tau$ relation.

\section{Discussion}
We have presented new measurements of the pairwise kSZ signal from the most recent ACTPol maps combined with the ACT data. We used the LSS DR11 catalog from the Baryon Oscillation Spectroscopic Survey to trace the positions of galaxy clusters and found a $4.1\sigma$ detection of the kSZ signal from the 20000 most luminous sources. By fitting to a pairwise velocity template corresponding to the same average redshift and mass cut of this sample we have found an average optical depth $\bar{\tau} = (1.46\pm0.36)\times 10^{-4}$. We have explored the dependence of the estimated signal-to-noise ratio on the method used to reconstruct the covariance matrix. The correlation between bins of comoving separation has a complicated structure and we show that the approach used to estimate the covariance matrix has important effects on the estimated significance of the measurement. The most stable and conservative method is based on simulations of the CMB sky plus noise. 

%This approach is more time consuming, but recovers the proper correlation and includes properly the contribution from CMB cosmic variance. The bootstrap method tends to underestimate the uncertainties especially for $r > 150$ Mpc, while the jackknife approach provides a S/N consistent with simulations but is less precise in describing the correlation between bins. We have also analyzed the dependence of the signal on the minimum luminosity of the sample and found that the signal-to-noise ratio increases when lowering the minimum luminosity, that is when including more sources, up to $L>7.9\times 10^{10} L_{\odot}$ for the DR11 LSS catalog. Including sources with lower luminosity does not increase the significance of the detection. While the error bars are reduced when including more sources, some systematic effects may become dominant; for instance these low luminosity galaxies may be satellites which do not trace the centers of clusters well. We measured the tSZ signal associated with the same sources used for the kSZ analysis. We reconstructed the Comptonization parameter from the tSZ temperature decrement in bins of luminosity and used results from hydrodynamical simulations to estimate the average optical depth per cluster. 
We found that the optical depth estimated with tSZ measurements is consistent with the one estimated by fitting the kSZ pairwise momentum measurements to the analytical pairwise velocity, assuming a $\Lambda$CDM cosmology. This shows that using tSZ data may be a viable approach for normalizing the mean pairwise velocity.

This work represents an extension and an improvement over the first kSZ measurement presented in \cite{2012PhRvL.109d1101H} using ACT data and a sample of the DR9 BOSS galaxies. We have used an aperture photometry filtering approach and a more conservative but more stable covariance matrix estimation, based on simulations of the CMB sky and noise. We have also confirmed the evidence of the kSZ effect in \cite{2012PhRvL.109d1101H} over the same range of comoving separations.

The recent work presented by Soergel et al.\cite{2016arXiv160303904S} reported a $4.2\sigma$ measurement of the pairwise momentum using CMB data from the South Pole Telescope and data from the Dark Energy Survey, using an approach similar to the one presented in this paper, fitting to a model template. The Soergel et al. work is particulary interesting because it uses photometric redshifts, showing that, with an appropriate treatment of the photometric uncertainty, photometric catalogs can provide significant evidence for the pairwise momentum. Soergel et al. used clusters with richness $\lambda_s>20$ and average redshift $\bar{z}=0.5$, while the SDSS-based redMaPPer catalog overlapping with the CMB map used in our work has a lower average redshift, $\bar{z}=0.35$. The best fit optical depth value reported in \cite{2016arXiv160303904S} is $(3.75\pm0.89)\times10^{-3}$, higher than the values reported in this paper. This difference is due to the more massive clusters used in the Soergel et al. analysis ($M_{500}>10^{14} M_{\odot}$). Moreover, the matched filter approach used in their work is more sensitive to the signal at the center of the clusters, while the aperture photometry used here measures the average optical depth in a wider area around the center of the clusters. The S/N reported by Soergel et al. is comparable to the one presented in this work, even though our analysis uses spectroscopic redshifts. Several factors may contribute to this, such as the use of a catalog of clusters at higher redshifts which might be less affected by centering issues, and the different approach used to estimate the covariance matrix (jackknife instead of simulations).

Different methods have been used and proposed to extract the kSZ signal. Schaan et al.\ in \cite{Schaan:2015uaa} used a velocity reconstruction approach to measure the amplitude of the kSZ signal as a function of the angular radius around the clusters, reporting a 2.9 and 3.3 $\sigma$ evidence of the kSZ signal, depending on the velocity reconstruction method used. The ACTPol CMB map used by Schaan et al. is similar to the one used in this work (about 660 deg$^2$) and it was combined with the CMASS galaxy catalog from BOSS DR10. Galaxy stellar mass estimates were converted to total masses for the host halos and then to an optical depth by using the cosmological baryon abundance. This method provides a potential probe for the free electron fraction in galaxy clusters and for the baryon profile of clusters. This different approach is complementary to pairwise kSZ measurements and could be used to understand and remove potential systematic effects.

Recently, Ferraro et al. \cite{Ferraro:2016ymw} and Hill et al. \cite{Hill:2016dta} have studied and applied a projected field approach, consisting of squaring the CMB anisotropy maps. Foreground-cleaned CMB temperature maps constructed from multi-frequency Planck and WMAP data were filtered, squared, and cross-correlated with galaxy measurements from the Wide-field Infrared Survey Explorer (WISE), finding 3.8-4.5$\sigma$ evidence for the kSZ, depending on the galaxy bias constraints. This method requires knowledge of the redshift distribution but not of the redshifts of single objects, allowing use of photometric data without treating redshift uncertainties and provides an additional method for probing the baryon distribution as a function
of scale and redshift.

Improved measurements from ACTPol are expected with the 2015 data, which has roughly 3 times larger overlap with BOSS \cite{fdb_spie}. Future surveys, like Advanced ACTPol \cite{Henderson:2015nzj}, SPT-3G \cite{Benson:2014qhw}, the Simons Observatory \footnote{\url{http://simonsobservatory.org/}} and a stage IV CMB experiment \cite{Abazajian:2013oma} can apply the kSZ pairwise statistic, velocity reconstruction and projected fields methods to achieve strong detections of the kSZ effect and to probe the baryon content of galaxy clusters. Multi-frequency data will enable measurements optical depth and peculiar velocities simultaneously and for single clusters. With these data the pairwise kSZ signal may become an important new cosmological probe that is complementary to current observables and is  able to constrain cosmology over a large range of physical scales.

\section{Acknowledgements}
We thank Shirley Ho, Daisuke Nagai, Eduardo Rozo, Eli Rykoff, Bjoern Soergel and Kyle Story for useful discussions. We thank the referee for the very useful comments which strengthened our paper. This work was supported by the U.S. National Science Foundation (NSF) through awards AST-0408698 and AST-0965625 for the ACT project, as well as awards PHY-0855887 and PHY-1214379. FDB and MDN acknowledge support from NSF grant AST-1454881 and AST-1517049. SA and AK were partly supported by NSF-AST-1312380. EMV acknowledges support from the NSF Graduate Research Fellowship Program under Grant No.DGE-1650441. NB acknowledges the support from the Lyman Spitzer Fellowship. RD thanks CONICYT for grants FONDECYT 1141113 and PIA Anillo ACT-1417. RH acknowledges funding from the Dunlap Institute.  Funding was also provided by Princeton University, the
University of Pennsylvania, Cornell University and a Canada Foundation for Innovation (CFI) award
to UBC. ACT operates in the Parque Astron\'omico Atacama in northern Chile
under the auspices of the Comisi\'on Nacional de Investigaci\'on Cient\'ifica y
Tecnol\'ogica de Chile (CONICYT). Computations were performed on the GPC
supercomputer at the SciNet HPC Consortium. SciNet is funded by the CFI under
the auspices of Compute Canada, the Government of Ontario, the Ontario Research
Fund -- Research Excellence; and the University of Toronto. 
Colleagues at RadioSky provide logistical support and keep operations in
Chile running smoothly. We also thank the Mishrahi Fund and the
Wilkinson Fund for their generous support of the project. The development of multichroic detectors and lenses was supported by
NASA grants NNX13AE56G and NNX14AB58G. %add: Soergel, Story,Daisuke Nagai, ...    

Funding for SDSS-III has been provided by the Alfred P. Sloan Foundation, the Participating Institutions, the National Science Foundation, and the U.S. Department of Energy Office of Science. The SDSS-III web site is http://www.sdss3.org/.

SDSS-III is managed by the Astrophysical Research Consortium for the Participating Institutions of the SDSS-III Collaboration including the University of Arizona, the Brazilian Participation Group, Brookhaven National Laboratory, Carnegie Mellon University, University of Florida, the French Participation Group, the German Participation Group, Harvard University, the Instituto de Astrofisica de Canarias, the Michigan State/Notre Dame/JINA Participation Group, Johns Hopkins University, Lawrence Berkeley National Laboratory, Max Planck Institute for Astrophysics, Max Planck Institute for Extraterrestrial Physics, New Mexico State University, New York University, Ohio State University, Pennsylvania State University, University of Portsmouth, Princeton University, the Spanish Participation Group, University of Tokyo, University of Utah, Vanderbilt University, University of Virginia, University of Washington, and Yale University.
% BIB
%%%%%%%%%%%%%%%%%%%%%%%%%%%%%%%%%%%%%%%%%%%%
\clearpage
\bibliographystyle{apsrev}
\bibliography{big_bib.bib}

\end{document}